# Create, run, share, publish, and reference your LC-MS, FIA-MS, GC-MS, and NMR data analysis workflows with the Workflow4Metabolomics 3.0 Galaxy online infrastructure for metabolomics

Yann Guitton[1,§], Marie Tremblay-Franco[2,§], Gildas Le Corguillé[3], Jean-François Martin[2], Mélanie Pétéra[4], Pierrick Roger-Mele[5], Alexis Delabrière[5], Sophie Goulitquer[6], Misharl Monsoor[3], Christophe Duperier[4], Cécile Canlet[2], Rémi Servien[2], Patrick Tardivel[2], Christophe Caron[7], Franck Giacomoni[4,*], and Etienne A. Thévenot[5,*]

[1]LUNAM Université, Oniris, Laboratoire d'Etude des Résidus et Contaminants dans les Aliments (LABERCA), Nantes, F-44307, France
[2]Toxalim (Research Centre in Food Toxicology), Université de Toulouse, INRA, ENVT, INP-Purpan, UPS, MetaboHUB, Toulouse, France
[3]UPMC, CNRS, FR2424, ABiMS, Station Biologique, 29680, Roscoff, France
[4]INRA, UMR 1019, PFEM, MetaboHUB, 63122, Saint Genes Champanelle, France
[5]CEA, LIST, Laboratory for Data Analysis and Systems' Intelligence, MetaboHUB, F-91191 Gif-sur-Yvette, France
[6]INSERM-UBO UMR1078-ECLA, IBSAM, Faculty of Medicine, University of Brest, 29200 Brest, France
[7]INRA, Ingenum, Toulouse, France

[§]These authors contributed equally to this work.

[*]**Corresponding authors:** contact@workflow4metabolomics.org

# Abstract

Metabolomics is a key approach in modern functional genomics and systems biology. Due to the complexity of metabolomics data, the variety of experimental designs, and the variety of existing bioinformatics tools, providing experimenters with a simple and efficient resource to conduct comprehensive and rigorous analysis of their data is of utmost importance. In 2014, we launched the Workflow4Metabolomics (W4M; http://workflow4metabolomics.org) online infrastructure for metabolomics built on the Galaxy environment, which offers user-friendly features to build and run data analysis workflows including preprocessing, statistical analysis, and annotation steps. Here we present the new W4M 3.0 release, which contains twice as many tools as the first version, and provides two features which are, to our knowledge, unique among online resources. First, data from the four major metabolomics technologies (i.e., LC-MS, FIA-MS, GC-MS, and NMR) can be analyzed on a single platform. By using three studies in human physiology, alga evolution, and animal toxicology, we demonstrate how the 40 available tools can be easily combined to address biological issues. Second, the full analysis (including the workflow, the parameter values, the input data and output results) can be referenced with a permanent digital object identifier (DOI). Publication of data analyses is of major importance for robust and reproducible science. Furthermore, the publicly shared workflows are of high-value for e-learning and training. The Workflow4Metabolomics 3.0 e-infrastructure thus not only offers a unique online environment for analysis of data from the main metabolomics technologies, but it is also the first reference repository for metabolomics workflows.

249 Words

**Graphical abstract**

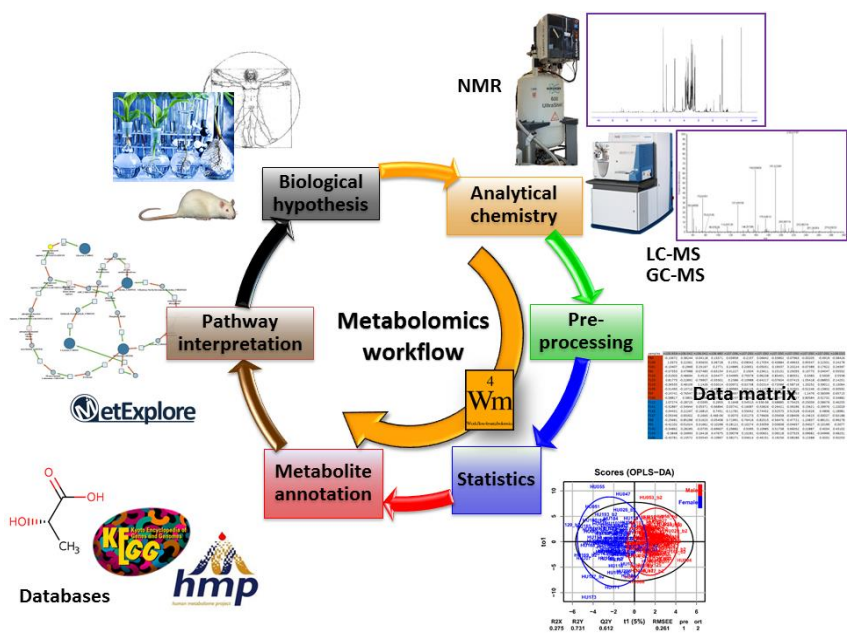

**Highlights**

- A single online resource for LC-MS, FIA-MS, GC-MS and NMR metabolomics data analysis
- 40 tools for data processing, statistical analysis, and metabolite identification
- The user-friendly Galaxy interface for building, running, saving and sharing workflows
- The first repository for the publication of workflows and histories with a permanent DOI
- Key materials and interactive environment for e-learning and teaching

**Abbreviations:**

BPA, Bisphenol A

DOI, Digital Object Identifier

e-infrastructure, online infrastructure

FIA, Flow Injection Analysis

GC, Gas Chromatography

HR, High-Resolution

LC, Liquid Chromatography

MS, Mass Spectrometry

NMR, Nuclear Magnetic Resonance

(O)PLS-DA, (Orthogonal) Partial Least Squares - Discriminant Analysis

PCA, Principal Component Analysis

ppm, parts-per-million

QC, Quality control sample (pool of all samples)

SVM, Support Vector Machine

VIP, Variable Importance in Projection

W4M, Workflow4Metabolomics

**Acknowledgements**

The Workflow4Metabolomics infrastructure is supported by the French Institute of Bioinformatics (IFB; ANR-11-INBS-0013) and by the French Infrastructure for Metabolomics and Fluxomics (MetaboHUB; ANR-11-INBS-0010). The authors thank Dominique Rolin, Florence Castelli and Jessica Wedow for critical reading of the manuscript, in addition to the reviewers for their valuable comments. We are also grateful to Simon Dittami for sharing the GC-MS data, and acknowledge the support from the IDEALG project (ANR-10-BTBR-04, "Investissements d'avenir, Biotechnologies-Bioressources").

# 1. Introduction

Metabolomics is the comprehensive quantification and characterization of the small molecules involved in metabolic chemical reactions (Oliver et al., 1998; Nicholson et al., 1999). It is a promising approach in functional genomics and systems biology for phenotype characterization and biomarker discovery, and has been applied to agriculture, biotechnology, microbiology, environment, nutrition, and health (Holmes et al., 2008; Chen et al., 2012; Rolin, 2013; Kell and Oliver, 2016; Johnson et al., 2016). Complementary analytical approaches, such as Nuclear Magnetic Resonance (NMR) or High-Resolution Mass Spectrometry (HRMS) coupled to Liquid Chromatography (LC) or Gas Chromatography (GC), can be used after minimal sample preparation. These technologies allow routine detection of hundreds to thousands of signals in a variety of biological samples such as cell cultures, organs, biofluids, or biopsies (Cuperlovic-Culf et al., 2010; Brown et al., 2012). Due to the high complexity and large amount of signals generated, however, data analysis remains a major challenge for high-throughput metabolomics (Johnson et al., 2015).

Analysis of metabolomics data (i.e., computational metabolomics) can be divided into three steps: preprocessing of raw data to generate the sample by variable matrix of intensities, statistical analysis to detect variables of interest and build prediction models, and annotation of variables to provide insight into their chemical and biological functions (Fig. 1). The two latter steps (statistics and annotation) can also be performed in the reverse order to get a first-pass overview of the dataset content by performing automatic query of metabolite databases. Furthermore, each step is subdivided into multiple successive, or alternative tasks. For example, preprocessing includes peak detection, denoising, and alignment. Statistical analysis involves normalization, univariate hypothesis testing and multivariate modeling. Finally, annotation relies on peak or spectrum matching with in-house and public databases of metabolites and spectra. This results in a high number of possible combinations of individual tasks to analyze in a dedicated data set. In addition, new methods and software tools constantly emerge to further expand, or refine, metabolomics analysis. Each of them has specific parameters and installation requirements. Typical data analysis by successive use of various software is time-consuming, repetitive, and error-prone: switching from one software to the other requires multiple steps of data manipulation (import/export, up/download, format conversion). Additionally, the workflow (i.e., the sequence of software tools and the parameter values) is not saved, thus preventing efficient and reproducible analysis.

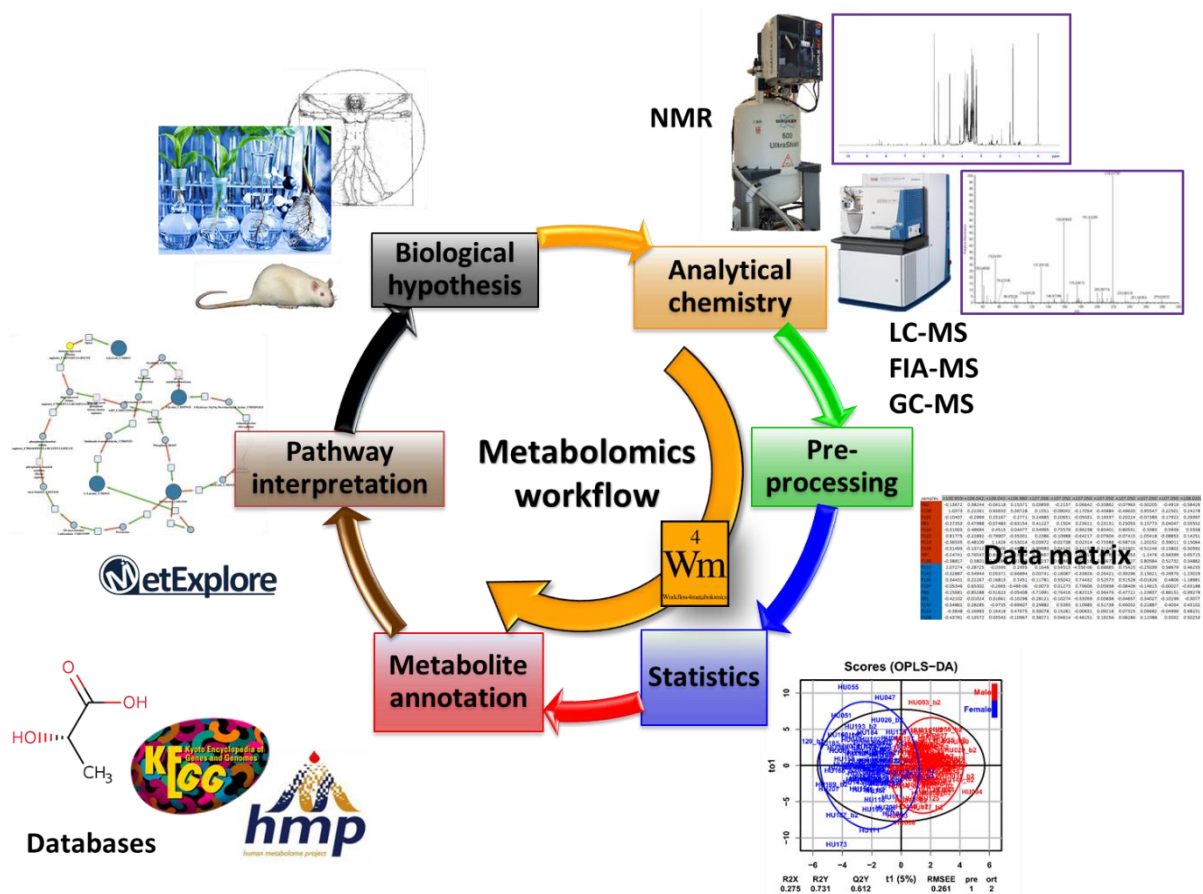

**Fig. 1.** The Workflow4Metabolomics 3.0 online infrastructure (e-infrastructure) for data analysis of metabolomics experiments. Metabolomics experiments start with the biological question to be addressed, which, in turn, defines the experimental design. Sample collection, preparation, and analysis with NMR or MS instruments are then performed with the appropriate quality controls (reagent blanks, sample pools, etc.). The preprocessing step generates the sample by variable matrix of peak intensities. Statistical analysis includes normalization and batch-effect correction, univariate hypothesis testing, multivariate modeling, and feature selection. Annotation relies on the query of compounds and spectral databases, such as KEGG (Kanehisa and Gotto, 2000) and HMDB (Wishart et al., 2007). Identified metabolites can then be linked within genome-scale reconstructed networks, such as MetExplore (Cottret et al., 2010). More than 40 modules (*tools*) are currently available on the Workflow4Metabolomics e-infrastructure for building comprehensive LC-MS, FIA-MS, GC-MS, and NMR preprocessing, statistical analysis, and metabolite annotation (Table 1).

Workflow management systems are software tools to compose and execute a series of computational tasks in a reproducible way (Leipzig, 2017). In the last decade, software environments with user-friendly features for creating, running, and sharing workflows have been developed, such as Galaxy (Giardine et al., 2005), Taverna (Hull et al., 2006), or KNIME (Berthold et al., 2006). Their graphical interfaces enable users who are not familiar with programming to build their workflow, by selecting tools and their parameters, and chain them in the desired order. Experimenters can therefore concentrate on the scientific design of the analysis and the interpretation of results, without worrying about software installation, command lines, scripts, data format and data management.

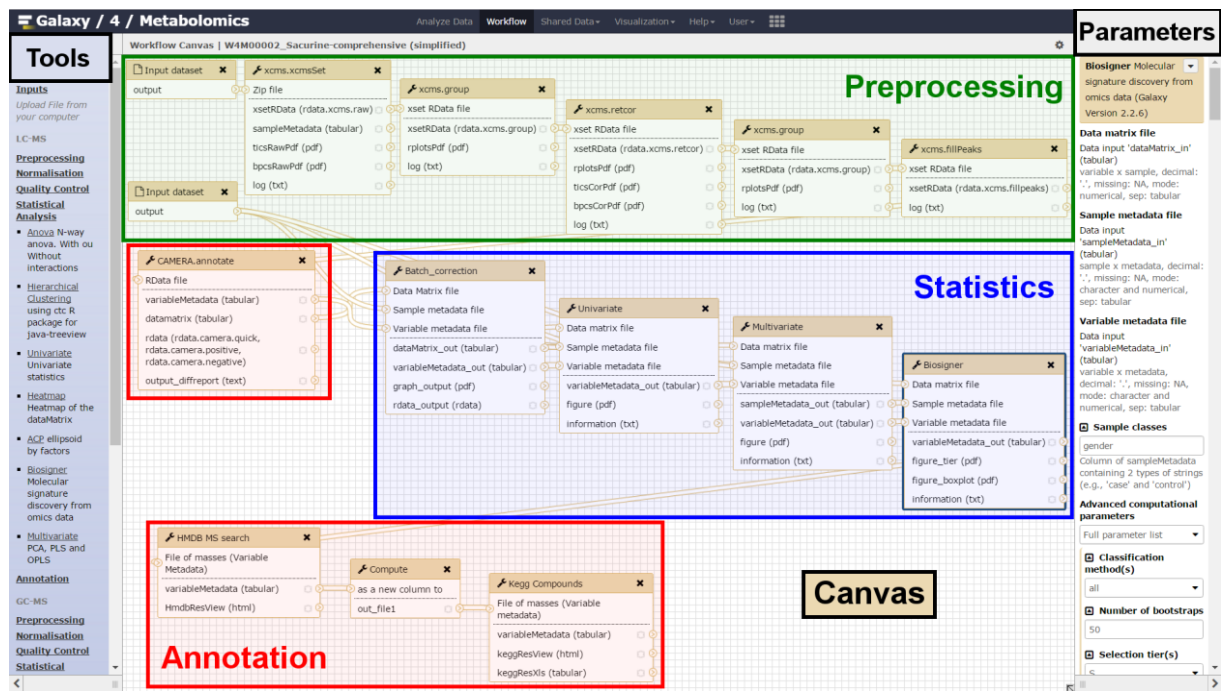

**Fig. 2.** Galaxy features to create workflows on the W4M e-infrastructure. Selected computational *tools* (left panel), with the specific parameters values (right panel), can be chained to build the *workflow* (middle panel). Alternatively, workflows can be directly extracted from the current *history* by selecting *Extract workflow* from the *History options* menu.

Galaxy is a major environment for workflow management available through a classic web-browser, with more than 50,000 users worldwide and hundreds of tools available (Boekel et al., 2015). The project started in the genomic community (Giardine et al., 2005; Goecks et al., 2010), and further expanded to other omics fields such as proteomics (Boekel et al., 2015; Jagtap et al., 2014; Jagtap et al., 2015). The Galaxy environment provides intuitive and powerful features that enable the experimenter to build and run complex workflows. For example, a tool can be re-run after changing a single parameter value with only two mouse clicks. The *workflow* (chained tools + selected parameter values) can be designed within a graphical editor interface (named *canvas*; Goecks et al., 2010; Fig. 2), and then run on the input data to obtain a *history* (workflow + attached input and output data). Alternatively, a history can be built by tuning the successive tools sequentially on the selected data set, and, once the history is completed, by extracting the corresponding workflow. A key feature is that both the workflow and the history can be saved and shared between users.

To develop Galaxy pipelines for metabolomics, and make them available to the user community worldwide, we created the Workflow4Metabolomics online infrastructure (W4M e-infrastructure; Giacomoni et al., 2015). At that time, W4M already provided 20 tools (some of them being newly developed for the infrastructure, while others corresponded to the integration of existing tools), which enabled the build of comprehensive workflows for LC-HRMS data analysis (Table 1). W4M not only gave access to the Galaxy environment to build workflows, but also offered a high-performance computing environment to run the analyses, online documentations, and a help-desk served by 8 bioinformaticians from the Core Team.

Here, we present the new 3.0 version of the W4M e-infrastructure. The 20 new tools (Table 1) enable advanced workflows not only for MS technologies (LC-MS, GC-MS and Flow Injection Analysis: FIA-MS) but also for NMR data. Furthermore, the complete histories can be referenced online with a permanent DOI, thus enabling fully reproducible analyses. To demonstrate how the new computational features from W4M 3.0 can be used to address biological issues, we selected three real LC-MS, GC-MS, and NMR case studies from published studies in human physiology, mouse toxicology, and algae evolution. In the next section (*Case Studies*), we briefly recall the objective of the three studies and the analytical methods used to generate the raw data. In the *Results* section, we then create three workflows to analyze the data, and compare the outputs with the published results.

**Table 1**

List of available tools on the W4M online infrastructure. The 20 new tools from the 3.0 release are indicated with (N). The source code of the W4M tools is available on the Galaxy toolshed (https://toolshed.g2.bx.psu.edu).

| Step | Workflow | | | | Tool | Description |
|---|---|---|---|---|---|---|
| Preprocessing | LCMS | | | | xcms.xcmsSet | Peak detection within each sample file |
| | LCMS | | | | xcms.xcmsSet Merger (N) | Merging a collection of xset.RData outputs before the grouping step |
| | LCMS | | | | xcms.group | Peak matching across samples |
| | LCMS | | | | xcms.retcor | Retention time correction |
| | LCMS | | | | xcms.fillpeaks | Imputation of missing intensities |
| | LCMS | | | | xcms.summary (N) | Summary of XCMS analysis |
| | LCMS | | | | CAMERA.annotate | Annotation of peak isotopes, adducts, and fragments |
| | LCMS | | | | CAMERA.combine (N) | Combining annotations from positive and negative ionization modes |
| | | FIAMS | | | proFIA (N) | Preprocessing of FIA-HRMS data |
| | | | GCMS | | metaMS.runGC (N) | GC-MS data preprocessing using metaMS package |
| | | | | NMR | NMR Alignment (N) | Spectra alignment based on the Cluster-based Peak Alignment (CluPA) algorithm |
| | | | | NMR | NMR Bucketing (N) | Bucketing and integration of spectra |
| Normalization Quality Control | LCMS | FIAMS | GCMS | | Determine batch correction | Choosing between linear and loess methods for batch correction |
| | LCMS | FIAMS | GCMS | | Batch correction | Corrects intensities for signal drift and batch-effects |
| | All | | | | Normalize (N) | Normalization of the dataMatrix |
| | All | | | | Transformation | Transforms the dataMatrix intensity values |
| | All | | | | Quality Metrics (N) | Metrics and graphics to check the quality of the data |
| | All | | | | Multilevel (N) | Extracts the within-subject variation from the dataMatrix in case of repeated measurements |
| Statistics | LCMS | | GCMS | | Metabolite Correlation Analysis | Filtering metabolites with correlated intensities |
| | All | | | | Univariate | Univariate statistics |
| | All | | | | Anova | N-way Anova with or without interactions |
| | All | | | | Multivariate | PCA, PLS(-DA) and OPLS(-DA) |
| | All | | | | Heatmap (N) | Heatmap of the dataMatrix |
| | All | | | | Hierarchical Clustering | Hierarchical clustering with export for Treeview visualization |
| | All | | | | Biosigner (N) | Molecular signature discovery from omics data |
| Annotation | LCMS | FIAMS | | | HR2 formula | Computes chemical formulas for (ionized) molecule masses |
| | LCMS | FIAMS | | | HMDB MS search | Searches the HMDB database by (ionized) molecule masses |
| | LCMS | FIAMS | | | Kegg Compounds | Searches the KEGG database by molecules masses |
| | LCMS | FIAMS | | | Lipidmaps | Searches the LIPID MAPS database by molecules masses |
| | LCMS | FIAMS | | | Chemspider | Searches the ChemSpider database by molecules masses |
| | LCMS | FIAMS | | | Bank in house (N) | Searches a local database by ion masses (and retention times) |
| | LCMS | FIAMS | | | LC/MS matching (N) | Searches a local database by ion masses (and retention times) |
| | LCMS | FIAMS | | | MassBank | Searches the MassBank database by molecules masses |
| | LCMS | | | | MassBank spectrum search (N) | Searches the MassBank spectral database by pseudo-spectra |
| | LCMS | | | | ProbMetab (N) | Refined annotation through incorporation of metadata and network information |
| | | | GCMS | | Golm Metabolome Database (N) | Searches the Golm Metabolome Database with spectra in the .msp format |
| | | | | NMR | NMR Annotation (N) | Annotation of complex spectra mixture and estimation of metabolite proportions |
| Data Handling | All | | | | Generic Filter | Removes samples or variables according to numerical or qualitative criteria |
| | All | | | | Table Merge (N) | Merging dataMatrix with a sampleMetadata or variableMetadata |
| | All | | | | Check Format (N) | Checks the formats of the dataMatrix, sampleMetadata, and variableMetadata files |

# 2. Case Studies

We have selected three case studies which illustrate the diversity of the biological issues, experimental designs, and analytical technologies in metabolomics. Here, we briefly describe the context, objective, and analytical methods of these three published studies.

## 2.1. Sacurine human physiological study (LC-HRMS)

Human urine has been used since Antiquity for disease prediction. Nowadays, this biofluid shows great promise for biomarker discovery in metabolomics. Characterization of the variations of the urine metabolome with age, BMI, and gender, is therefore critical not only to better understand human physiology, but also to avoid confounding effects in biomarker studies. Since physiological information about urine concentrations is scarce in metabolomics databases, a cohort of 184 volunteers from the CEA research institute was studied (Roux et al., 2012; Thevenot et al., 2015). Urine samples were analyzed by Ultra-high Performance Liquid Chromatography (Hypersil GOLD C18 column) coupled to High-Resolution Mass Spectrometry (LTQ-Orbitrap Discovery, Thermo Fisher Scientific).

The work by Roux et al. (2012) focused on the identification of the metabolites in urine, while the second study (Thevenot et al., 2015) analyzed the variations of their concentrations with age, body mass index (BMI), and gender. Raw data were preprocessed with XCMS (Smith et al., 2006) and annotated with CAMERA (Kuhl et al., 2012), and a selection of metabolites of putative interest were identified at levels 1 and 2 (Metabolomics Standard Initiative; Sumner et al., 2007) by matching with the KEGG (Kanehisa and Gotto, 2000), HMDB (Wishart et al., 2007) and METLIN (Smith et al., 2005) databases, followed by interpretation of additional MS/MS fragmentation experiments (Roux et al., 2012). Quantification of their intensities was refined by visual determination of the peak limits in the raw data (Quan Browser tool from the Xcalibur software; Thevenot et al., 2015).

The raw files, both in the proprietary (RAW; profile) and in the open (mzML; centroid) formats, are publicly available (MTBLS404) from the MetaboLights repository (Haug et al., 2013).

## 2.2. *Ectocarpus* (brown algae) ecological study (GC-MS)

The brown alga *Ectocarpus* (Ectocarpales, Phaeophyceae) is a rare example of freshwater colonization by a marine species. To understand the molecular mechanisms involved in the transition between these habitats, a freshwater strain (FWS, accession CCAP 1310/196, origin Hopkins River Falls, Victoria, Australia) was exposed to either medium (1.6 parts per thousand of medium, ppt) or seawater (32 ppt) salinity, and compared with the genome-sequenced Sea Water strain (SWS, accession CCAP 1310/4, origin San Juan de Marcona, Peru) cultured in seawater medium only (Dittami et al., 2012).

Metabolites profiles were analyzed by Gas Chromatography coupled to Mass Spectrometry (GC-MS): algal samples were first harvested by filtration 2 h after the beginning of the light phase; samples (50 mg) were then dried with a paper towel and ground in liquid nitrogen; finally, algal powder was extracted with methanol, and ribitol (200 µM) was added as an internal standard. The complete extraction procedure, derivatization steps, and chromatographic conditions are described in Dittami et al. (2011).

## 2.3. Bisphenol A *Mus musculus* toxicological study (NMR)

The aim of this study was to assess the effect of perinatal exposure to Bisphenol A (BPA) on the brain metabolome (Cabaton et al., 2013). Brain samples were collected on 21-days-old male mice, whose mothers were exposed to either 0.025 or 0.25 µg BPA/kg body weight/day during gestation and lactation. Twenty four aqueous extracts (11 extracts from mice exposed to 0.025 µg and 13 from mice exposed to 0.25 µg) were analyzed on a Bruker DRX-600-Avance NMR spectrometer operating at 600.13 MHz for $^1$H resonance frequency using an inverse detection 5 mm $^1$H-$^{13}$C-$^{15}$N cryoprobe attached to a CryoPlatform (the preamplifier cooling unit). The $^1$H NMR spectra were acquired at 300 K using the Carr-Purcell-Meiboom-Gill (CPMG) spin-echo pulse sequence with pre-saturation, with a total spin-echo delay (2nτ) of 100 ms to attenuate broad signals from proteins and lipoproteins. A total of 128 transients were collected resulting in 32,000 data points using a spectral width of 12 ppm, a relaxation delay of 2.5 s and an acquisition time of 2.28 s. Prior to Fourier Transformation, an exponential line broadening function of 0.3 Hz was applied to the FID. All NMR spectra were phased and baseline corrected with the TopSpin software (Bruker).

# 3. Results

The Workflow4Metabolomics online infrastructure (W4M e-infrastructure) offers a user-friendly and computationally efficient environment to build, run, and share workflows (see the *Workflow Management with W4M* section in the supplementary material). The W4M 3.0 release provides a total of 40 tools, from preprocessing, through statistical analysis, and up to annotation, including 20 new modules for advanced LC-HRMS analysis, FIA-MS, GC-MS and NMR workflows (Table 1).

## 3.1. Analysis of LC-MS, GC-MS, and NMR metabolomics data sets on W4M

To demonstrate how the tools can be combined on W4M to address biological issues, we designed three computational workflows to analyze data sets from published studies in human (Thevenot et al., 2015), algae (Dittami et al., 2012), and mouse (Cabaton et al., 2013).

### 3.1.1. Physiological variations of the human urine metabolome: The *W4M00001_Sacurine-statistics* and *W4M00002_Sacurine-comprehensive* LC-HRMS histories

To characterize the physiological variations of the human urine metabolome with age, body mass index (BMI), and gender, samples from a cohort of adult volunteers from the Saclay research institute (hence the "sac[lay]urine" name) have been analyzed by LC-HRMS (Thevenot et al., 2015).

To investigate how the statistical analysis from the published study could be performed on W4M, the *Sacurine-statistics* workflow was first built with the following steps (Fig. 3 and statistical steps from the supplementary Table S1):

1. **Batch correction**: Correction of the signal drift by local regression (*loess*) modeling of the intensity trend in pool samples (Dunn et al., 2011); Adjustment of offset differences between the two analytical batches by using the average of the pool intensities in each batch (van der Kloet et al., 2009)
2. **Quality Metrics**: Variable quality control by discarding features with a coefficient of variation above 30% in pool samples
3. **Normalization**: Intensity normalization by sample osmolality
4. **Transformation**: Log10 transformation
5. **Quality Metrics**: Sample outlier filtering by using three statistics: Weighted Hotellings'T2 distance (Tenenhaus et al., 1999), Z-score of one of the intensity distribution deciles (Alonso et al., 2011), and Z-score of the number of missing values (Alonso et al., 2011)
6. **Univariate**: Non-parametric univariate hypothesis testing of Spearman correlation with age or BMI, and of difference between gender medians
7. **Multivariate**: Multivariate modeling by Orthogonal Partial Least Squares (OPLS) of the age, BMI, and gender responses
8. **Heatmap**: Visualization of sample and variable clusters (by using the 1 - *cor* dissimilarity, where *cor* is the Spearman correlation)

The workflow consists of 18 tools. Importantly, several of these computational methods are available online only on W4M. In addition, the statistical tools are generic (except the first one, **Batch correction**, which is focused on LC-MS data), and can thus be applied to any omic data set.

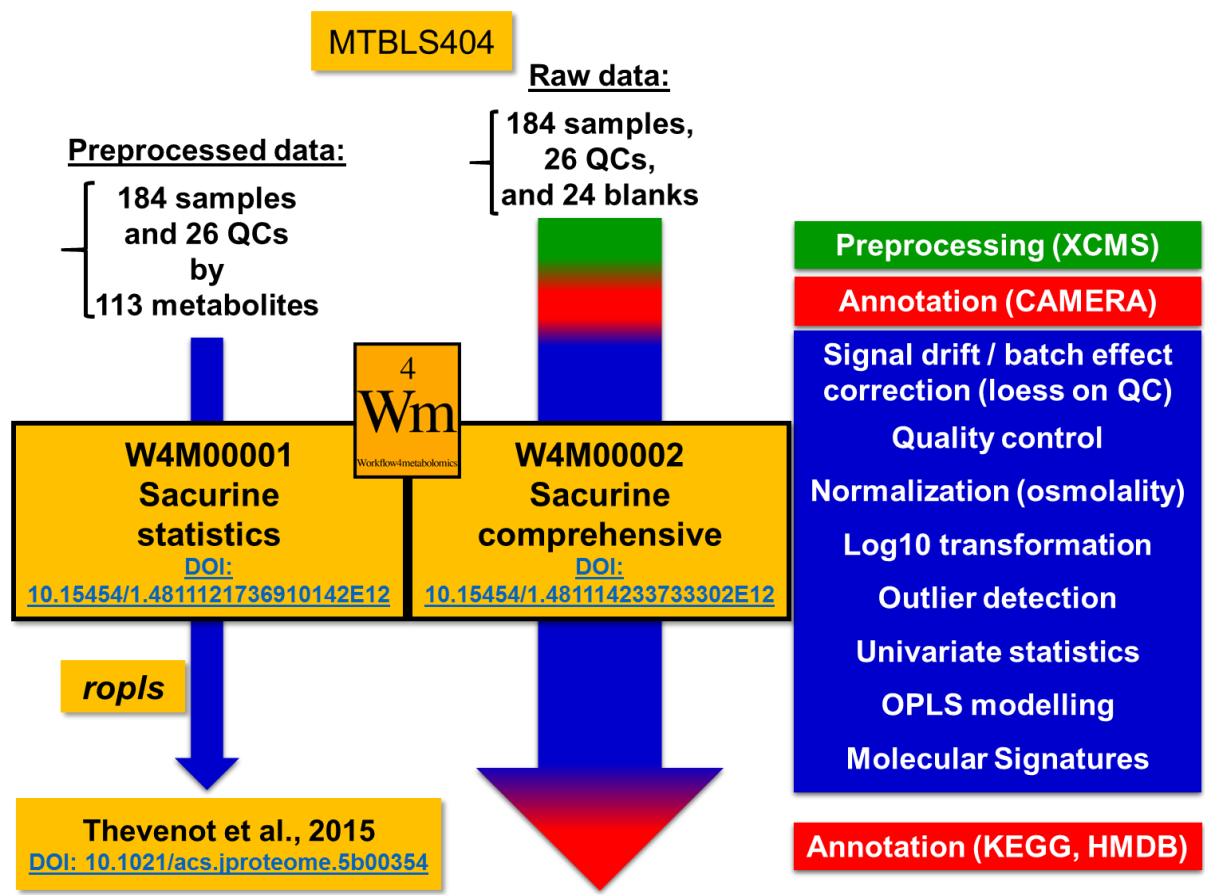

**Fig. 3.** Overview of the *W4M00001 Sacurine-statistics* and *W4M00002 Sacurine-comprehensive* workflows (DOI:10.15454/1.4811121736910142E12 and DOI:10.15454/1.481114233733302E12). The input data (184 samples, 26 QC pools, and 24 reagent blanks) were acquired in the negative ionization mode, in two batches (117 samples each). Note that the same workflows can be applied to the data from the positive ionization mode, except that the polarity parameter must be switched in the annotation tools (i.e., CAMERA, KEGG, and HMDB). The raw data are publicly available in the MetaboLights repository (MTBLS404).

The workflow was applied to the preprocessed table of peak intensities from 210 samples (184 urine of volunteers + 26 injection of the QC pool) and 113 identified metabolites from the negative ionization mode (see the *Case Studies* section for the description of the study, and the *Workflow Management with W4M* section for the formats of the dataMatrix.tsv, sampleMetadata.tsv, and variableMetadata.tsv input files). The running time on W4M was only a few minutes and the resulting history contained 53 files (total size: 4.0 MB). Statistical results were identical to those from the publication by Thevenot et al. (2015). In particular, the peak table used for univariate and multivariate statistics contained 183 samples and 109 metabolites, since the quality control steps discarded the HU_096 sample (*p*-values of the Hotelling's T2 and decile Z-score < $10^{-4}$) and 4 variables (coefficient of variation in QC injections > 30%). This peak table is also identical to the data set included in the *ropls* Bioconductor package (Thevenot et al., 2015). The *Sacurine-statistics* history was publicly shared on the W4M e-infrastructure. It was assigned the *W4M00001* reference ID and the permanent [DOI:10.15454/1.4811121736910142E12](DOI:10.15454/1.4811121736910142E12) Digital Object Identifier (DOI) link for permanent access and citation (see the *Referencing Histories* section below).

To perform the full analysis of the *sacurine* dataset (i.e., not restricting the analysis to the statistical steps nor to the identified metabolites only), we designed the *Sacurine-comprehensive* workflow, starting with the preprocessing of the raw files, then performing the statistical analysis and finally annotating the full peak table (Fig. 3A):

1. ***xcms.xcmsSet***, ***xcms.group***, ***xcms.retcor***, and ***xcms.fillPeaks***: Preprocessing of the raw files with XCMS (Smith et al., 2006)
2. ***CAMERA.annotate***: Annotation of isotopes, adducts and fragments with CAMERA (Kuhl et al., 2012)
3. Statistical analysis (identical to the *Sacurine-statistics* workflow, with the addition of the ***Biosigner*** tool)
4. ***HMDB MS search*** and ***Kegg Compounds***: Annotation by m/z matching to the HMDB (Wishart et al., 2007) and KEGG (Kanehisa and Goto, 2000) metabolite databases;

The *Sacurine-comprehensive* workflow includes 29 tools (21 being unique; Table S1) and provides a comprehensive example of the analysis of LC-HRMS metabolomics data (Fig. 3). In the statistical section, the ***Biosigner*** tool (Rinaudo et al., 2016) was included to further identify significant molecular signatures for classification between genders (see below).

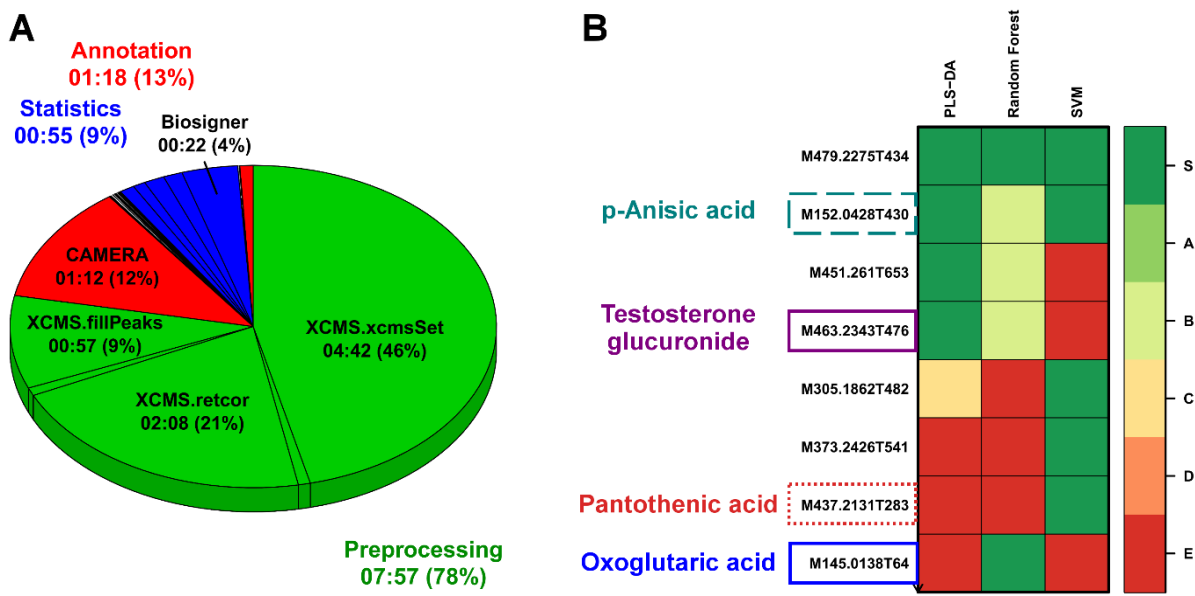

**Fig. 4.** Results from the *W4M00002_sacurine-comprehensive* history (DOI:10.15454/1.481114233733302E12). (A) Computation time of the full analysis. Only tools running for more than 15 min are labeled. The total running time of the whole history (29 tools) was 10:10 hours. (B) Significant metabolite signatures for the discrimination between genders by each of the three PLS-DA, Random Forest, and Support Vector Machine (SVM) classifiers were selected independently with the **Biosigner** tool (Rinaudo et al., 2016). The *biosigner* algorithm iteratively selects only the features whose random permutation in the test subset decrease the prediction performances (Rinaudo et al., 2016). The metabolites from the final signature (i.e., those which passed all the selection iterations) are indicated in the corresponding classifier column with a dark green color. Metabolites which were discarded during one of the previous rounds are indicated with a color gradient from light green to dark red (by decreasing number of selection rounds). Dashed (respectively, dotted) borders around metabolite names indicate the $^{13}$C isotope (respectively, the Pantothenic acid dimer).

The 234 raw files (184 samples + 26 injections of the Quality Control pool + 24 blanks) in mzML format (centroid mode) were split into two subfolders (one "blank" directory containing the blank reagent files only, and one "bio" directory with the remaining 210 human and pool samples), before uploading as a single zipped file (17.6 GB). Application of the *Sacurine-comprehensive* workflow to this dataset resulted in a history containing 94 files, after a running time of about 10 hours (Fig. 4A). The majority of the time was spent in preprocessing with the **XCMS** tools and, to a lesser extent, in annotation with **CAMERA.annotate**, as these algorithms work on the raw files. In the statistical section, the **Biosigner** tool required 22 min of computation time, due to the large number of models built during the internal bootstrapping approach (Fig. 4A).

The preprocessing step with the **XCMS** tools and the original parameter values (Roux et al., 2012) generated a peak table (*dataMatrix*) containing 4,667 features (Table S1), including 342 ions previously described in urine (Roux et al., 2012). These results were comparable with the 4,114 features (including 334 characterized ions) from the original peak table obtained at the time of the publication by Roux et al. (2012). We further evaluated the *centWave* algorithm which has been shown to provide better peak detection on high-resolution data sets compared with the original *matchedFilter* method (Tautenhahn et al., 2008). Such a comparison is straightforward on W4M due to both features provided by Galaxy for modifying only a few parameter values (in the **xcms.xcmsSet** tool) and re-running the workflow, and the high-performance environment of the infrastructure. A higher number of variables of interest (372 of the previously characterized metabolites in urine) were detected by *centWave*, with the parameter values suggested for an UPLC/Orbitrap acquisition system (Patti et al., 2012). The total size of the peak table after the quality control filtering steps (which discard the ions with concentrations less than twice the concentration in blank reagent samples, or with a coefficient of variation in pool samples superior to 30% after signal drift correction) was also higher (3,120 features), including 85% of the *matchedFilter* ions. The *centWave* method was therefore selected for the reference workflow and history described hereafter (Table S1).

As the identification of molecular signatures for prediction is of high interest in biomarker studies, the **Biosigner** tool was added to the statistical section of the workflow. **Biosigner** selects features that are significant for the predictive performance of either a PLS-DA, a Random Forest, or a Support Vector Machine classifier (Rinaudo et al., 2016). Eight features were selected in at least one of the molecular signatures, including one $^{13}$C isotopes and one dimer (Fig. 4B). The PLS-DA, Random Forest, and Support Vector Machine trained on each selected signatures (containing less than 0.2 % of the initial variables) achieved high classification performances (84%, 90%, and 92%, respectively). Three variables, which were annotated by the KEGG query and also matched the characterized metabolites described in urine (Roux et al., 2012), had been selected previously by **Biosigner** on the restricted dataset from the *W4M00001_Sacurine-statistics* history (Rinaudo et al., 2016): Testosterone glucuronide, p-Anisic Acid (or 4-Methoxybenzoic acid), and Oxoglutaric acid (or alpha-Ketoglutaric acid). This confirms that results could be reproduced on W4M using a comprehensive workflow starting from the raw files. It also highlights 4 additional ions providing high classification performance (including M479.2275T434) that were not previously described in the urine metabolome.

### 3.1.2. Evolutionary mechanisms involved in the adaptation of marine species to freshwater: The *W4M00004 GCMS Algae* history

Colonization of freshwater by marine organisms, as observed in *Ectocarpus* (brown algae), is a rare event. To understand the mechanisms involved in such an adaptation, a Fresh Water Strain (FWS) cultured in either low or seawater salinity, and a Sea Water Strain (SWS), were studied at several molecular levels, including metabolomics by GC-MS analysis (Dittami et al., 2012). At the time of the publication by Dittami and colleagues, no GC-MS tool was available on W4M. To reproduce the study results, we therefore implemented two new Galaxy tools for preprocessing and annotation (since the statistical tools already available on W4M can be applied to data from all technologies). First, the **metaMS.runGC** tool based on the *metaMS* package from Wehrens et al. (2014) was integrated for preprocessing. The main parameter is *FWHM* (full width at half maximum of chromatographic peaks), which is used for peak picking by the internal call to the *matchedFilter* algorithm from *XCMS* (Smith et al., 2006). The default value is 5 s, and can be modified within the *User_defined* setting. The **metaMS.runGC** function further groups individual ions within the same retention time window and same chromatographic profile into *pseudospectra*, by using the *groupFWHM* function from *CAMERA* (Kuhl et al., 2012). Each pseudospectrum usually (but not always) contains ions originating from a single compound (Wehrens et al., 2014). Finally, annotation information is added by matching pseudospectra against each other (or, optionally, against a database of standards with the *Use Personal Database* option). A table of retention index (RI) of each alkane used in the GC-MS experiment can also be provided (*Use RI* option): with that option, the RI information is added into the *variableMetadata.tsv* outputs, which can facilitate peak annotation. The main output of **metaMS.runGC** is a data matrix of pseudospectra intensities (sum of the intensities of all ions from the pseudospectrum), which can be used for downstream statistical analysis. It is important to note that, in this data matrix, each row corresponds to one compound, in contrast to the classical XCMS output where each row is a single ion feature. This is particularly interesting because unlike electrospray ionization commonly used in LC-MS, electron impact ionization in GC-MS generates a lot of ions for each compound.

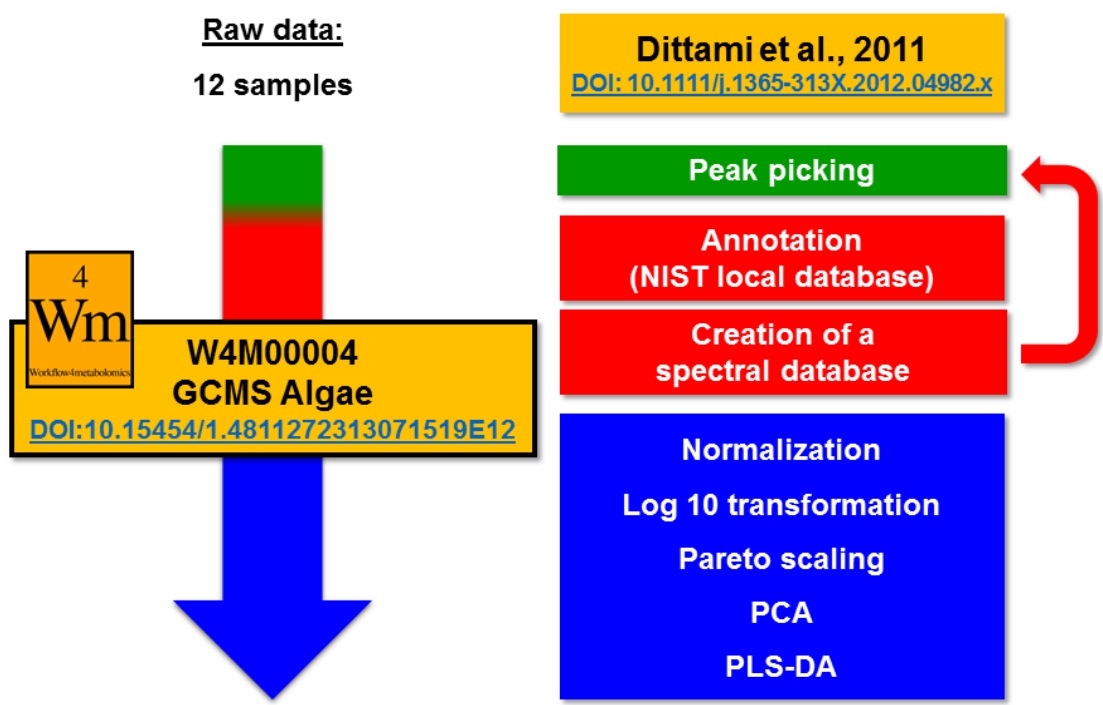

**Fig. 5.** *W4M00004 GCMS Algae* reference workflow (DOI:10.15454/1.4811272313071519E12). Twelve *Ectocarpus* samples (brown algae), corresponding to 4 biological replicates of 3 algal cultures (a freshwater strain in low and seawater saline conditions, and a marine strain as a control) were analyzed with an Agilent GC-MS instrument. Preprocessing, statistics, and annotation of the data set were performed with the ***metaMS.runGC***, ***Multivariate***, and ***Golm Metabolome Database*** tools, in addition to a local query of the NIST database (Table S2). An in-house spectral database was created and used to refine the peak picking (red arrow). Raw files (in the open NetCDF format) are publicly available on W4M (http://workflow4metabolomics.org/datasets).

Second, we developed a tool to search the ***Golm Metabolome Database*** (Kopka et al., 2005). This annotation tool uses the *peakspectra.msp* file generated by ***metaMS.runGC***. The *.msp* file can be further exported to query the commercial NIST database with the *mssearch* software (usually installed on GC-MS instruments). The resulting *.msp* file can be used as an in-house database (Wehrens et al., 2014) to annotate pseudospectra from future experiments.

To perform the analysis of the GC-MS data from the three *Ectocarpus* cultures (a freshwater strain in low and seawater saline conditions, and a marine strain as a control), the *W4M00004 GCMS Algae* workflow was then designed (Fig. 5 and Table S2):

1. Preprocessing with **metaMS.runGC** (the default parameter values were used)
2. Annotation by using the **Golm Metabolome Database** tool and the local NIST database
3. Creation of a spectral database with identified pseudospectra
4. **metaMS.runGC**: Reprocessing using the in-house database for refined pseudospectra alignment
5. **Normalization**: Intensity normalization by dry weight of sample
6. **Multivariate**: Exploratory data analysis with PCA
7. **Multivariate**: PLS-DA modeling of the three distinct algal cultures

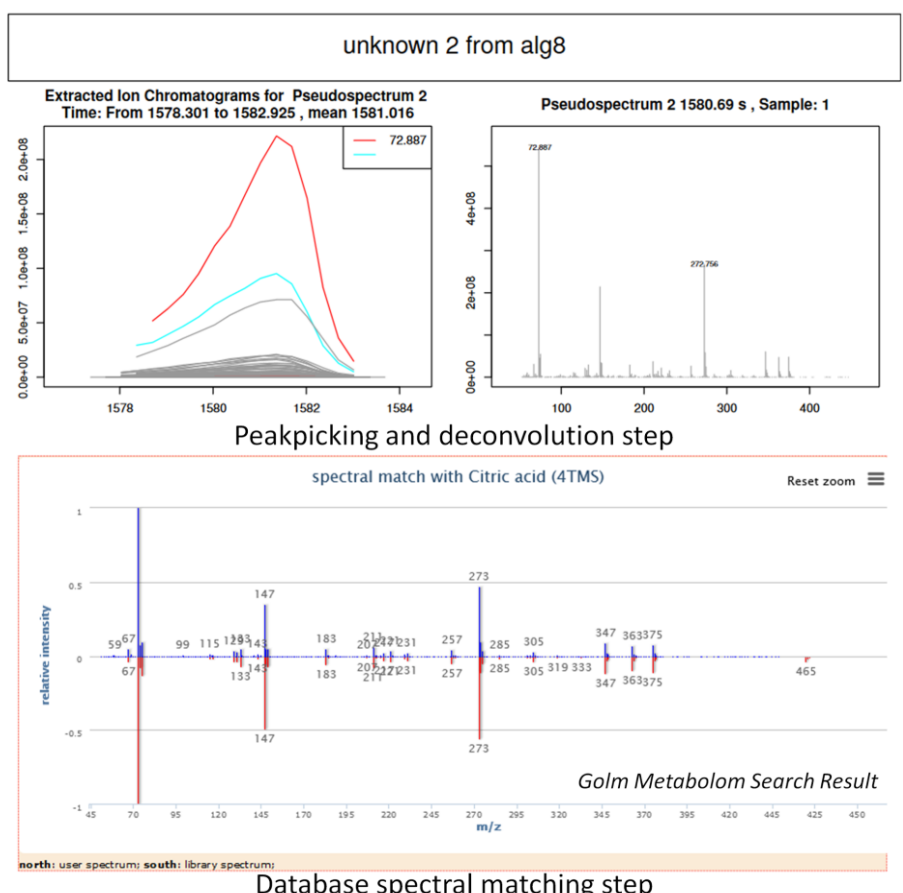

**Fig. 6.** Quality control and annotation of the "Unknown 2" pseudospectrum. Top: The EIC of the pseudospectrum clearly indicates that all grouped ions belong to a single compound (no co-elution). Only ions with correlated chromatographic profiles are grouped by the metaMS algorithm, resulting in a cleaned pseudospectrum for further spectral database annotation (Wehrens et al., 2014). Bottom: After matching to the Golm and NIST databases, the "Unknown 2" pseudospectrum was annotated as citric acid, which is the correct identification (confirmed by injection of the pure compound).

Prior to data upload, the 12 GC-MS raw files (4 biological replicates of the 3 cultures) were converted from the Agilent commercial format (*.D) into the open NetCDF format, by using the Agilent Chemstation software ('export as AIA/ANDI Files' menu; Note that the freely available ProteoWizard software can also be used for file conversion; Chambers et al., 2012). The converted files from the W4M00004 study are available for download from W4M as a unique zipped file (260.8 MB; http://workflow4metabolomics.org/datasets).

Application of the workflow to the raw data resulted in a history containing 21 files generated in 11 minutes (7 min for peak picking and 4 min for statistics). The data matrix contained 52 pseudospectra. The quality of each pseudospectrum was checked visually on the *GCMS_EIC.pdf* output (Fig. 6, top). All pseudospectra were then matched against the Golm and NIST spectral databases, by using the *peakspectra.msp* file (Fig. 6, bottom). As a control, the internal standard ribitol was confirmed to be the "Unknown 5". Importantly, "Unknown 2" and "Unknown 4" were annotated as citric acid and mannitol, respectively (Fig. 6, bottom). Surprisingly, mannitol was not detected in two samples (alg 2 and alg 3), because the retention time shift (15 s) was superior to the default threshold in **metaMS.runGC** (3 s). We therefore used an option from **metaMS.runGC** which enables to refine the alignment of pseudospectra between samples by providing a database of reference spectra (Wehrens et al., 2014): first, we created a spectral database with ribitol, citric acid, and mannitol, from the *peakspectra.msp* file generated previously; second, we re-run the **metaMS.runGC** tool with this additional spectral information, and successfully detected mannitol in all samples (feedback step on Fig. 5). We could then perform the downstream statistical analyses. The matrix of intensities (*dataMatrix*) was first normalized by dry weight of sample. Exploratory data analysis was then performed by PCA: distinct clusters corresponding to three algal cultures were observed on the score plot. In particular, a clear metabolic shift was observed for the FWS cultures from the low versus seawater saline conditions. The decreased concentrations of mannitol, which is a putative osmoprotector, in the FS strain, were in accordance with the results from Dittami et al. (2012). Finally, PLS-DA discrimination between the 3 cultures resulted in a classifier with a significant ($p < 0.05$, when compared with models built after random permutation of the labels) and high Q2 value (Q2Y = 0.84). These multivariate analyses, which were not included in the original publication, therefore provide complementary information.

### 3.1.3. Brain toxicity of Bisphenol A: The *W4M00006 NMR BPAMmusculus* history

The aim of this study was to assess the effect on the brain metabolome of perinatal exposure to Bisphenol A (BPA), an endocrine disruptor widely used in plastics and resins (Cabaton et al. 2013). To analyze NMR data on W4M, specific preprocessing and annotation modules have been developed for the 3.0 version (see below). The implemented workflow (Fig. 7) corresponds to a complementary pairwise comparison study (BPA0.025 vs BPA0.25), which was not presented in the original publication. The 0.25 and 0.025 µg BPA/kg body weight/day treatments correspond to 1/100 and 1/000 of the tolerable daily intake (TDI: "Estimated maximum amount of an agent, expressed on a body mass basis, to which individuals may be exposed daily over their lifetimes without appreciable health risk"; World Health Organization, 2004) and were picked up to demonstrate that even at very low doses of exposure, BPA is still modulating differently the brain metabolome of the CD1 mice.

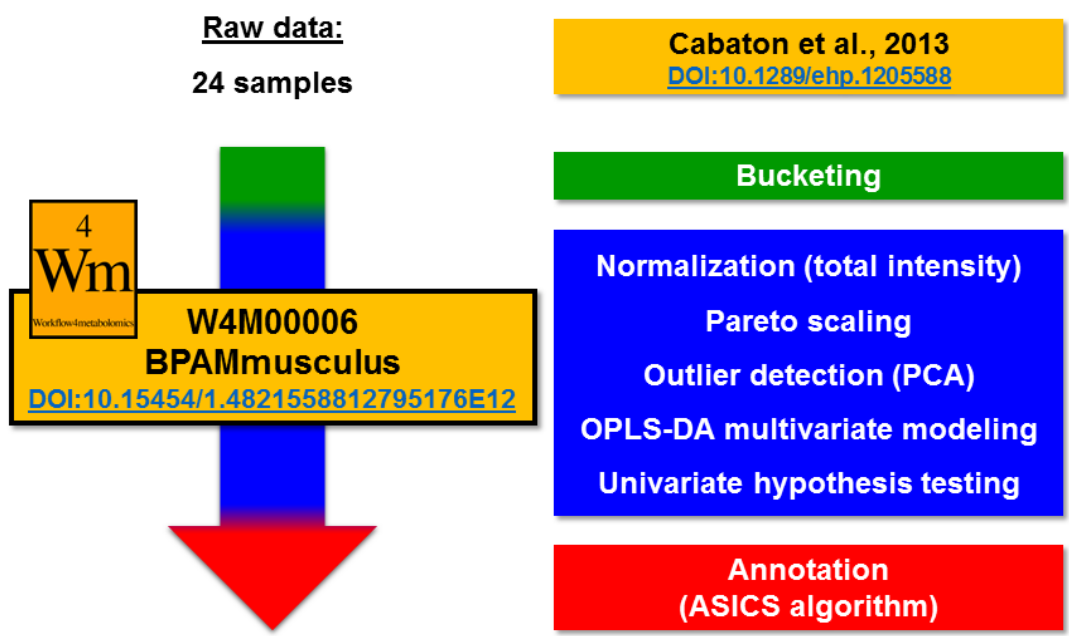

**Fig. 7.** *W4M00006 BPAMmusculus* reference workflow (DOI:10.15454/1.4821558812795176E12). The input data (24 samples) were acquired using $^1$H NMR spectroscopy. TopSpin (Bruker) preprocessed files (Bruker format) are available on W4M for download (http://workflow4metabolomics.org/datasets).

The 7 tools from the *BPAMmusculus* workflow are detailed in Table S3 and illustrated on Fig. 8. First, the 24 TopSpin-preprocessed spectra were uploaded into W4M as a single zipped file (see the *Workflow Management with W4M* section). Second, spectra were segmented into 809 buckets by using the **NMR Bucketing** tool: this tool divides the whole spectrum into "small" fixed-size windows (e.g., 0.01 ppm). In addition, spectrum regions corresponding to water, solvent or contaminant resonances can be excluded. Finally, the sum of intensities inside each bucket (area under the curve) is computed by using the trapezoidal method. Third, spectra were normalized to the Total Intensity with the **NMR Normalization** tool. The objective of sample normalization is to make the data from all samples directly comparable with each other (i.e., to remove systematic biological and technical variations). The **NMR Normalization** tool includes 3 normalization methods: Total intensity (each bucket integration is divided by the integration of the total spectrum), quantitative variable (e.g., sample weight, osmolality), and Probabilistic Quotient Normalization (PQN; Dieterle et al., 2006), where each spectrum is compared to a reference sample (e.g., the median spectrum of control samples). Fourth, exploratory data analysis and multivariate modeling were performed with the *Multivariate* tool. PCA was first used to detect outliers: two observations were excluded for subsequent analyses. Then, an OPLS-DA classifier of the two treatment doses was built: the model was significant (permutation test *p*-value < 0.05), and 157 variables (buckets) having a VIP value > 0.8 were selected. In parallel, univariate analysis of differences between the two doses was performed with the Wilcoxon-Mann-Whitney test (*Univariate* tool). No significant difference was observed after correction for multiple testing (False Discovery Rate set to 5%).

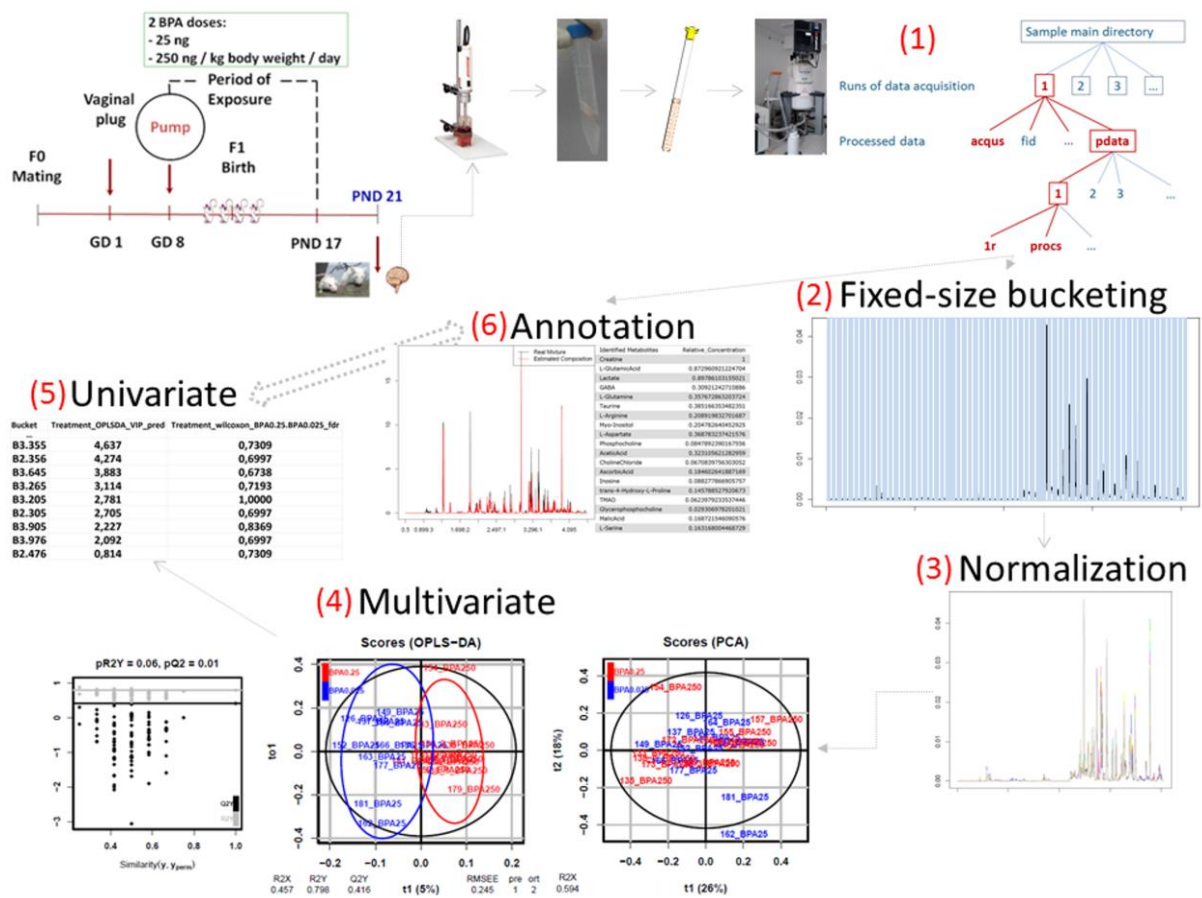

**Fig. 8.** *W4M00006 BPAMmusculus*: From experiment to metabolite annotation. Top: Experimental design, sample preparation, and acquisition of NMR spectra (Bruker files). (1-3) Data reduction (bucketing + normalization). (4) Multivariate analysis (PCA and OPLS-DA score plots, and permutation tests). (5-6) Annotation of significant metabolites.

Sixth, buckets were annotated with the *NMR Annotation* tool. This tool decomposes any input spectrum from a complex biological matrix into a mixture of spectra from pure compounds provided as a reference database. The internal database currently contains 175 spectra of pure compounds, which were acquired on a Bruker Avance III 600 MHz NMR spectrometer, at pH 7.0. The deconvolution algorithm uses a penalized regression to compute a parsimonious list of non-zero coefficients (i.e., proportions) for the reference spectra detected in the mixture (Tardivel et al., submitted). The output *proportionEstimation* table contains the identified metabolites with their estimated relative concentration (the highest concentration being arbitrarily set to 1). The **NMR Annotation** tool was applied to one spectrum from each of the two classes (BPA0.025 and BPA0.25), and resulted in the annotation of 39 metabolites. Among them, the Glutamic Acid and the GABA neurotransmitters had high VIP predictive values (4.1 for the 2.35 ppm bucket, and 2.6 for the 2.3 ppm bucket, respectively), which confirmed the previous results showing a decrease of the concentrations of these metabolites following exposure to BPA (Cabaton et al., 2013). Furthermore, these new analyses showed that two other neurotransmitters, Taurine (3.44-3.41, 3.26-3.24 ppm) and Aspartate (3.91-3.89, 2.7-2.69 ppm), had high discriminative values (VIP of 2.9 for the 3.26 ppm bucket, and 2.2 for the 3.90 ppm bucket, respectively). To our knowledge, this is the first time that comprehensive NMR workflows (from preprocessing to identification) can be created, run, and published online.

## 3.2. Referencing histories

The *histories* from each case study (i.e., workflow and the associated input data and output result files) were further *published* on W4M (i.e., shared with the community) and referenced with a digital object identifier (DOI) which can be cited in publications (Table 2). Referenced histories can be imported by any user into his/her account, and the workflow can be extracted from the history with the *Extract Workflow* functionality, e.g., for application to new data sets.

Making workflows and associated data available to the community is essential to demonstrate the value and the reproducibility of the analysis (Mons et al., 2011). As for raw data, journal editors will increasingly recommend that the process of generating the results (code, parameter values, output data) is made available on reference repositories. Funding agencies such as European Programs also require that the generated data are made public on reference online resources. Finally, sharing analyses gives experimenters the opportunity to receive feedback on their results, get cited, and initiate new collaborations. While databases for raw data and metabolites already exist, the W4M e-infrastructure is the first repository for data analysis workflows dedicated to metabolomics.

Referencing histories on W4M is straightforward (see the "Referenced Workflows and Histories" section on the home page). Authentication is required to access the shared histories in Galaxy. Extra anonymous credentials for reviewers, however, may be provided to authors from reference histories when submitting their manuscript. Six histories have already been referenced (Table 2), and have been cited in publications such as Thevenot et al. (2015), Rinaudo et al. (2016), and Peng et al. (2017).

**Table 2**

Publicly referenced histories (workflows and associated data and result files; http://workflow4metabolomics.org/referenced_W4M_histories). The DOI points to a landing page which details the main steps, data size, and name of the maintainer, and gives access the whole history online.

| ID | Raw data | Technol. | Species | Matrix | Factor(s) | Nb samples | Size | Publication |
|---|---|---|---|---|---|---|---|---|
| W4M00001_ Sacurine-statistics DOI:10.15454/1.4811121736910142E12 | MTBLS404 | LC-MS | H. sapiens | urine | age, BMI, gender | 210 | 4 MB | Thevenot et al., 2015 DOI:10.1021/acs.jproteome.5b00354 |
| W4M00002_ Sacurine-comprehensive DOI:10.15454/1.481114233733302E12 | MTBLS404 | LC-MS | H. sapiens | urine | age, BMI, gender | 234 | 18 GB | Thevenot et al., 2015 DOI:10.1021/acs.jproteome.5b00354 |
| W4M00003_ Diaplasma DOI:10.15454/1.4811165052113186E12 | N/A | LC-MS | H. sapiens | plasma | diabetic type | 63 | 11 MB | Rinaudo et al., 2016 DOI:10.3389/fmolb.2016.00026 |
| W4M00004_ GCMS-Algae DOI:10.15454/1.4811272313071519E12 | N/A | GC-MS | E. siliculosus | alga | salinity | 12 | 260 MB | Dittami et al., 2012 DOI:10.1111/j.1365-313X.2012.04982.x |
| W4M00005_ Ractopamine-Pig DOI:10.1545 | MTBLS384 | LC-MS | S. Scrofa | serum | Ractopamine | 164 | 327 MB | Peng et al., 2017 DOI:10.1007/s11306-017-1212-0 |
| 4/1.4811287270056958E12 | | | | | | | | |
| W4M00006_BPA-Mmusculus DOI:10.15454/1.4821558812795176E12 | W4M | NMR | Mus Musculus | brain | BPA dose | 24 | 7 MB | Cabaton et al., 2013 DOI:10.1289/ehp.1205588 |

# 4. Discussion

High-throughput analysis of the metabolic phenotype has a profound impact on the understanding of biochemical reactions and physiology, and prediction of disease (Peyraud et al., 2011; Balog et al., 2013; Etalo et al., 2015; Li et al., 2016; Weiss et al., 2016). To cope with the volume and complexity of data generated by modern high-resolution MS and NMR instruments, adapt to the diversity of experimental designs, and ensure rigorous and reproducible analyses, there is a strong need for user-friendly, modular, and computationally efficient software platforms. To demonstrate how the Workflow4Metabolomics 3.0 e-infrastructure meets this workflow challenge, we designed, run, and referenced online three comprehensive analyses of previously published LC-MS, GC-MS, and NMR data studies from human physiology, mouse toxicology, and alga evolution. The workflow sequence, parameter values, and critical points are detailed to provide examples of critical design and interpretation of analyses. The outputs of W4M analyses confirmed the published results: for each case study, key metabolites were selected and annotated, with significant concentration differences either between gender, BMI, and age (human study), or after perinatal exposition to Bisphenol A (mouse study), or in response to salinity stress (algae study). Furthermore, they shed new lights on the datasets, e.g., by suggesting variations of neurotransmitter concentrations even at the lowest doses of BPA exposure (mouse study). Workflow building and running was very efficient: the *Sacurine-comprehensive* workflow, which contains 29 tools for preprocessing, statistical analysis and annotation, could be run on 234 raw files in a few hours. Workflow management was straightforward, as illustrated by the comparison between the *matchedFilter* and *centWave* approaches for LC-MS preprocessing. Together, these results demonstrate that comprehensive LC-MS, GC-MS, and NMR data analyses can be readily designed and run on the W4M e-infrastructure.

The number of available tools on W4M 3.0 for pre-processing, statistical analysis and annotation, 40, has doubled since the previous release (Giacomoni et al., 2015), and now allows to analyze LC-MS, FIA-MS, GC-MS, and NMR data. Importantly, several tools implement original methods from the Core Team which provide unique features to the W4M workflows, such as the **Biosigner** tool for selection of significant molecular signatures for PLS-DA, Random Forest, or SVM classifiers, the **NMR Annotation** tool for annotation of NMR spectra, or the ***proFIA*** tool for the preprocessing of data from Flow Injection Analysis coupled to High-Resolution Mass Spectrometry (FIA-HRMS; Delabriere et al., *under review*). Furthermore, to cope with the computer intensive preprocessing of LC-MS data, the ***xcms.xcmsSet*** and ***CAMERA.annotate*** tools can now be run in parallel (see the supplementary material). All W4M tools are implemented by a large core team of bioinformaticians and biostatisticians based on five metabolomics facilities, and supported in the long term by two national infrastructures, namely the French Institute of Bioinformatics (IFB; French Elixir node) and the National Infrastructure for Metabolomics and Fluxomics (MetaboHUB). In addition to ensuring a sustainability and high-performance computing environment, these large clusters provide cutting-edge technologies, know-how, and scientific expertise from both the experimental and computational fields. In the near future, new unique tools will be available on W4M (e.g., to extend the NMR workflow and to analyze MS/MS data). Moreover, complementary Galaxy tools have been recently described in metabolomics (e.g., for Direct Infusion MS data; Davidson et al., 2016), but also in complementary omics communities (Boekel et al., 2015) such as proteomics (Jagtap et al., 2015; Jagtap et al., 2014; Fan et al., 2015). Due to the modularity of the Galaxy environment, and the relative ease of wrapping existing code into Galaxy tools, the number of W4M tools and contributors should continue to expand rapidly. To help developers integrating their tools, an updated virtual machine and the code of the Galaxy modules are publicly available on the W4M GitHub (https://github.com/workflow4metabolomics) and the Galaxy Toolshed (https://toolshed.g2.bx.psu.edu/; '[W4M]' tag) repositories.

Workflow4Metabolomics brings the workflow management features of the Galaxy environment to the metabolomics user community through its online infrastructure. Online availability has many advantages compared with local installation: no local computing resources are needed, local software installation and update is not required, and the infrastructure can be directly accessed from anywhere. Two online platforms have recently emerged for LC-MS processing and annotation (XCMS Online; Tautenhahn et al., 2012), and statistical analysis (MetaboAnalyst; Xia et al., 2009), respectively. In contrast, W4M provides for the first time a single resource for the comprehensive analysis of either LC-MS, FIA-MS, GC-MS, or NMR metabolomics data. In addition, by building on the Galaxy environment, W4M provides the user with unique features to build, run, and share workflows and histories (e.g., with remote collaborators in multi-center or transdisciplinary projects).

In particular, W4M 3.0 now offers to reference a history publicly, by assigning a unique ID and DOI permanent link, which can be cited in publications. To our knowledge, this is the first time that workflows (and associated data) can be referenced. Furthermore, since the source code of the W4M tools is also publicly available (as discussed above), referenced analyses can be fully dissected and reproduced online (or locally) by the scientific community. The W4M infrastructure thus fills a gap between existing repositories for raw data, such as MetaboLights (Haug et al., 2013) or the Metabolomics Workbench (Sud et al., 2016), and the spectral and metabolite databases such as KEGG (Kanehisa and Goto, 2000), HMDB (Wishart et al., 2007), ChEBI (Degtyarenko et al., 2008) or MassBank (Horai et al., 2010). Further interoperability between the W4M workflow resource and the MetaboLights data repository is ongoing within the PhenoMeNal European consortium. In the open data era (Leonelli et al., 2013), the need for peer-reproduced workflows (Gonzalez-Beltran et al., 2015) and workflow storage (Belhajjame et al., 2015) is pivotal for good science and robust transfer to the clinic (Baker, 2005). Besides, funding agencies and journal editors already require data to be made publicly available (the MetaboLights repository is already recommended by scientific journals such as Metabolomics, the EMBO Journal, and Nature Scientific Data; Kale et al., 2016). W4M should therefore become the reference repository for metabolomics workflows.

To help users cope with data analysis concepts, parameter tuning, and critical interpretation of diagnostics and results, training is of major importance (Via et al., 2013; Weber et al., 2015). On the W4M infrastructure, remote e-learning is possible through many tutorials (http://workflow4metabolomics.org/howto). In addition, the reference histories provide detailed examples of workflows, and of table and figure outputs (http://workflow4metabolomics.org/referenced_W4M_histories). Furthermore, "hands-on" sessions using W4M can be readily organized since only an internet connection is needed to access the infrastructure (for users wishing to use W4M for training, please contact us at contact@workflow4metabolomics.org). Based on our experience, optimal results are achieved when users analyze their own data. We therefore regularly organize one-week courses combining practical presentations in the mornings, and tutoring sessions in the afternoons (Workflow4Experimenters, W4E; http://workflow4metabolomics.org/events). Such trainings with about 25 participants offer unique opportunities to discuss the designs, methods, and tools for comprehensive and rigorous data analysis.

In conclusion, the Workflow4Metabolomics 3.0 e-infrastructure provides experimenters with unique features to learn, design, run, share, and reference comprehensive LC-MS, FIA-MS, GC-MS, and NMR metabolomics data analyses.

# 5. References

Alonso A, Julia A, Beltran A, Vinaixa M, Diaz M, Ibanez L, Correig X, Marsal S. AStream: an R package for annotating LC/MS metabolomic data. Bioinformatics 2011;27:1339-1340.

Baker M. In biomarkers we trust? Nature Biotechnology 2005;23:297-304.

Balog J, Sasi-Szabo L, Kinross J, Lewis MR, Muirhead LJ, Veselkov K, Mirnezami R, Dezso B, Damjanovich L, Darzi A, Nicholson JK, Takats Z. Intraoperative Tissue Identification Using Rapid Evaporative Ionization Mass Spectrometry. Science Translational Medicine 2013;5:194ra93.

Belhajjame K, Zhao J, Garijo D, Gamble M, Hettne K, Palma R, Mina E, Corcho O, Gomez-Perez JM, Bechhofer S, Klyne G and Goble C. Using a suite of ontologies for preserving workflow-centric research objects. Web Semantics: Science, Services and Agents on the World Wide Web 2015;32:16-42.

Berthold MR, Cebron N, Dill F, Di Fatta G, Gabriel TR, Georg F, Meinl T, Ohl P, Sieb C, Wiswedel B. KNIME: the Konstanz Information Miner. In: Proceedings 4th Annual Industrial Simulation Conference (ISC), Workshop on Multi-Agent Systems and Simulation 2006.

Boekel J, Chilton JM, Cooke IR, Horvatovich PL, Jagtap PD, Kall L, Lehtio J, Lukasse P, Moerland PD, Griffin TJ. Multi-omic data analysis using Galaxy. Nature Biotechnology 2015;33:137-139.

Brown MV, McDunn JE, Gunst PR, Smith EM, Milburn MV, Troyer DA, Lawton KA. Cancer detection and biopsy classification using concurrent histopathological and metabolomic analysis of core biopsies. Genome Medicine 2012;4:33.

Cabaton NJ, Canlet C, Wadia PR, Tremblay-Franco M, Gautier R, Molina J, Sonnenschein C, Cravedi JP, Rubin BS, Soto A, Zalko D. Effects of low doses of Bisphenol A on the metabolome of perinatally exposed CD-1 mice. Environmental Health Perspectives 2013;121: 586-593.

Chambers MC, Maclean B, Burke R, Amodei D, Ruderman DL, Neumann S, Gatto L, Fischer B, Pratt B, Egertson J, Hoff K, Kessner D, Tasman N, Shulman N, Frewen B, Baker TA, Brusniak MY, Paulse C, Creasy D, Flashner L, Kani K, Moulding C, Seymour SL, Nuwaysir LM, Lefebvre B, Kuhlmann F, Roark J, Rainer P, Detlev S, Hemenway T, Huhmer A, Langridge J, Connolly B, Chadick T, Holly K, Eckels J, Deutsch EW, Moritz RL, Katz JE, Agus DB, MacCoss M, Tabb DL, Mallick P. A cross-platform toolkit for mass spectrometry and proteomics. Nature Biotechnology 2012;30:918-920.

Chen R, Mias G, Pook-Than JL, Jiang L, Lam H, Chen R, Miriami E, Karczewski K, Hariharan M, Dewey F, Cheng Y, Clark M, Im H, Habegger L, Balasubramanian S, O'Huallachain M, Dudley J, Hillenmeyer S, Haraksingh R, Sharon D, Euskirchen G, Lacroute P, Bettinger K, Boyle A, Kasowski M, Grubert F, Seki S, Garcia M, Whirl-Carrillo M, Gallardo M, Blasco M, Greenberg P, Snyder P, Klein T, Altman R, Butte AJ, Ashley E, Gerstein M, Nadeau K, Tang H, Snyder M. Personal omics profiling reveals dynamic molecular and medical phenotypes. Cell 2012;148:1293-1307.

Cottret L, Wildridge D, Vinson F, Barrett MP, Charles H, Sagot MF, Jourdan F. MetExplore: a web server to link metabolomic experiments and genome-scale metabolic networks. Nucleic Acids Research 2010;38:W132-W137.

Cuperlovic-Culf M, Barnett DA, Culf AS, Chute I. Cell culture metabolomics: applications and future directions. Drug Discovery Today 2010;15:610-621.

Davidson R, Weber R, Liu H, Sharma-Oates A, Viant M. Galaxy-M: A Galaxy workflow for processing and analyzing direct infusion and liquid chromatography mass spectrometry-based metabolomics data. GigaScience 2016;5:1-9.

Degtyarenko K, de Matos P, Ennis M, Hastings J, Zbinden M, McNaught A, Alcantara R, Darsow M, Guedj M, Ashburner M. ChEBI: a database and ontology for chemical entities of biological interest. Nucleic Acids Research 2008;36:D344-D350.

Delabriere A, Hohenester U, Junot C, Thevenot EA. proFIA: preprocessing data from flow injection analysis coupled to high-resolution mass spectrometry. *under review*.

Dieterle F, Ross A, Schlotterbeck G, Senn H. Probabilistic Quotient Normalization as Robust Method to Account for Dilution of Complex Biological Mixtures. Application in 1H NMR Metabonomics. Analytical Chemistry 2006;78:4281-4290.

Dittami SM, Gravot A, Renault D, Goulitquer S, Eggert A, Bouchereau A, Boyen C, Tonon T. Plant, Cell & Environment 2011;34:629-642.

Dittami SM, Gravot A, Goulitquer S, Rousvoal S, Peters AF, Bouchereau A, Boyen C, Tonon T. Towards deciphering dynamic changes and evolutionary mechanisms involved in the adaptation to low salinities in Ectocarpus (brown algae). The Plant Journal 2012;71:366-377.

Dunn WB, Broadhurst D, Begley P, Zelena E, Francis-McIntyre S, Anderson N, Brown M, Knowles JD, Halsall A, Haselden JN, Nicholls AW, Wilson ID, Kell DB,

Goodacre R (2011). Procedures for large-scale metabolic profiling of serum and plasma using gas chromatography and liquid chromatography coupled to mass spectrometry. Nature Protocols 2011;6:1060-1083.

Etalo DW, De Vos RC, Joosten MH, Hall RD. Spatially Resolved Plant Metabolomics: Some Potentials and Limitations of Laser-Ablation Electrospray Ionization Mass Spectrometry Metabolite Imaging. Plant Physiology 2015;169:1424-1435.

Fan J, Saha S, Barker G, Heesom KJ, Ghali F, Jones AR, Matthews DA, Bessant C. Galaxy integrated omics: web-based standards-compliant workflows for proteomics informed by transcriptomics. Molecular & Cellular Proteomics 2015;14:3087-3093.

Giacomoni F, Le Corguille G, Monsoor M, Landi M, Pericard P, Petera M, Duperier C, Tremblay-Franco M, Martin J, Jacob D, Goulitquer S, Thevenot E, Caron C. Workflow4Metabolomics: a collaborative research infrastructure for computational metabolomics. Bioinformatics 2015;31:1493-1495.

Giardine B, Riemer C, Hardison RC, Burhans R, Elnitski L, Shah P, Zhang Y, Blankenberg D, Albert I, Taylor J, Miller W, Kent WJ, Nekrutenko A. Galaxy: A platform for interactive large-scale genome analysis. Genome Research 2005;15:1451-1455.

Goecks J, Nekrutenko A, Taylor J, Team TG. Galaxy: a comprehensive approach for supporting accessible, reproducible, and transparent computational research in the life sciences. Genome Biology 2010;11:R86.

Gonzalez-Beltran A, Li P, Zhao J, Avila-Garcia MS, Roos M, Thompson M, van der Horst E, Kaliyaperumal R, Luo R, Lee TL, Lam TW, Edmunds SC, Sansone SA, Rocca-Serra P. From peer-reviewed to peer-reproduced in scholarly publishing: The complementary roles of data models and workflows in bioinformatics. PLoS ONE 2015;10:e0127612.

Haug K, Salek RM, Conesa P, Hastings J, de Matos P, Rijnbeek M, Mahendraker T, Williams M, Neumann S, Rocca-Serra P, Maguire E, Gonzalez-Beltran A, Sansone SA, Griffin JL, Steinbeck C. MetaboLights: an open-access general-purpose repository for metabolomics studies and associated meta-data. Nucleic Acids Research 2013;41:D781-D786.

Holmes E, Loo RL, Stamler J, Bictash M, Yap IKS, Chan Q, Ebbels T, De Iorio M, Brown IJ, Veselkov KA, Daviglus ML, Kesteloot H, Ueshima H, Zhao L, Nicholson JK, Elliott P. Human metabolic phenotype diversity and its association with diet and blood pressure. Nature 2008;453:396-400.

Horai H, Arita M, Kanaya S, Nihei Y, Ikeda T, Suwa K, Ojima Y, Tanaka K, Tanaka S, Aoshima K, Oda Y, Kakazu Y, Kusano M, Tohge T, Matsuda F, Sawada Y, Hirai MY, Nakanishi H, Ikeda K, Akimoto N, Maoka T, Takahashi H, Ara T, Sakurai N, Suzuki H, Shibata D, Neumann S, Iida T, Tanaka K, Funatsu K, Matsuura F, Soga T, Taguchi R, Saito K, Nishioka T. MassBank: a public repository for sharing mass spectral data for life sciences. Journal of Mass Spectrometry 2010;45:703-714.

Hull D, Wolstencroft K, Stevens R, Goble C, Pocock MR, Li P, Oinn T. Taverna: a tool for building and running workflows of services. Nucleic Acids Research 2006;34:W729-W732.

Jagtap PD, Johnson JE, Onsongo G, Sadler FW, Murray K, Wang Y, Shenykman GM, Bandhakavi S, Smith LM, Griffin TJ. Flexible and accessible Workflows for Improved Proteogenomic Analysis Using the Galaxy Framework. Journal of Proteome Research 2014;13:5898-5908.

Jagtap PD, Blakely A, Murray K, Stewart S, Kooren J, Johnson JE, Rhodus NL, Rudney J, Griffin TJ. Metaproteomic analysis using the Galaxy framework. Proteomics 2015;15:3553-3565.

Johnson CH, Ivanisevic J, Benton HP, Siuzdak G. Bioinformatics: The next frontier of metabolomics. Analytical Chemistry 2015;87:147-156.

Johnson CH, Ivanisevic J, Siuzdak G. Metabolomics: beyond biomarkers and towards mechanisms. Nature Reviews Molecular Cell Biology 2016;17:451-459.

Kale NS, Haug K, Conesa P, Jayseelan K, Moreno P, Rocca-Serra P, Nainala VC, Spicer RA, Williams M, Li X, Salek RM, Griffin JL, Steinbeck C. MetaboLights: An Open-Access Database Repository for Metabolomics Data. Current Protocols in Bioinformatics 2016;14.13.1-14.13.18.

Kanehisa M, Goto S. KEGG: Kyoto Encyclopedia of Genes and Genomes. Nucleic Acids Research 2000;28:27-30.

Kell DB, Oliver SG. The metabolome 18 years on: a concept comes of age. Metabolomics 2016;12:148.

van der Kloet FM, Bobeldijk I, Verheij ER, Jellema RH. Analytical error reduction using single point calibration for accurate and precise metabolomic phenotyping. Journal of Proteome Research 2009;8:5132-5141.

Kopka J, Schauer N, Krueger S, Birkemeyer C, Usadel B, Bergmuller E, Dormann P, Weckwerth W, Gibon Y, Stitt M, Willmitzer L, Fernie AR, Steinhauser D.

GMD@CSB.DB: the Golm Metabolome Database. Bioinformatics 2005;21:1635-1638.

Kuhl C, Tautenhahn R, Bottcher C, Larson TR, Neumann S. CAMERA: An integrated strategy for compound spectra extraction and annotation of Liquid Chromatography/Mass Spectrometry Data Sets. Analytical Chemistry 2012;84:283-289.

Leipzig J. A review of bioinformatic pipeline frameworks. Briefings in Bioinformatics 2017;18:530-536.

Leonelli S, Smirnoff N, Moore J, Cook C, Bastow R. Making open data work for plant scientists. Journal of Experimental Botany 2013;64:4109-4117.

Li D, Heiling S, Baldwin IT, Gaquerel E. Illuminating a plant's tissue-specific metabolic diversity using computational metabolomics and information theory. Proceedings of the National Academy of Sciences 2016;113:E7610-E7618.

Nicholson J, Lindon J, Holmes E. 'Metabonomics': understanding the metabolic responses of living systems to pathophysiological stimuli via multivariate statistical analysis of biological NMR spectroscopic data. Xenobiotica 1999;29:1181-1189.

Mons B, van Haagen H, Chichester C, Hoen PB, den Dunnen JT, van Ommen G, van Mulligen E, Singh B, Hooft R, Roos M, Hammond J, Kiesel B, Giardine B, Velterop J, Groth P, Schultes E. The value of data. Nature Genetics 2011;43:281-283.

Oliver SG, Winson MK, Kell DB, Baganz F. Systematic functional analysis of the yeast genome . Trends in Biotechnology 1998;16:373-378.

Patti GJ, Tautenhahn R, Siuzdak G. Meta-analysis of untargeted metabolomic data from multiple profiling experiments. Nature Protocols 2012;7:508-516.

Peng T, Royer AL, Guitton Y, Le Bizec B, Dervilly-Pinel G. Serum-based metabolomics characterization of pigs treated with ractopamine. Metabolomics 2017;13:77-91.

Peyraud R, Schneider K, Kiefer P, Massou S, Vorholt J, Portais JC. Genome-scale reconstruction and system level investigation of the metabolic network of Methylobacterium extorquens AM1. BMC Systems Biology 2011;5:189.

Rinaudo P, Boudah S, Junot C, Thevenot E. *biosigner*: A new method for the discovery of significant molecular signatures from omics data. Frontiers in Molecular Biosciences 2016;3.

Rolin D. Metabolomics Coming of Age with its Technological Diversity. Advances in Botanical Research 2013:67;2-693.

Roux A, Xu Y, Heilier JF, Olivier MF, Ezan E, Tabet JC, Junot C. Annotation of the human adult urinary metabolome and metabolite identification using ultra high performance liquid chromatography coupled to a linear quadrupole ion trap-orbitrap mass spectrometer. Analytical Chemistry 2012;84:6429-6437.

Smith CA, Maille GO, Want EJ, Qin C, Trauger SA, Brandon TR, Custodio DE, Abagyan R, Siuzdak G. METLIN: A Metabolite Mass Spectral Database. Therapeutic Drug Monitoring 2005;27:747-751.

Smith CA, Want EJ, O'Maille G, Abagyan R, Siuzdak G. XCMS: Processing mass spectrometry data for metabolite profiling using nonlinear peak alignment, matching, and identification. Analytical Chemistry 2006;78:779-787.

Sud M, Fahy E, Cotter D, Azam K, Vadivelu I, Burant C, Edison A, Fiehn O, Higashi R, Nair KS, Sumner S, Subramaniam S. Metabolomics Workbench: An international repository for metabolomics data and metadata, metabolite standards, protocols, tutorials and training, and analysis tools. Nucleic Acids Research 2016;44:D463-D470.

Sumner L, Amberg A, Barrett D, Beale M, Beger R, Daykin C, Fan T, Fiehn O, Goodacre R, Griffin J, Hankemeier T, Hardy N, Harnly J, Higashi R, Kopka J, Lane A, Lindon J, Marriott P, Nicholls A, Reily M, Thaden J, Viant M. Proposed minimum reporting standards for chemical analysis. Metabolomics 2007;3:211-221.

Tardivel PJC, Servien R, Concordet D. Familywise Error Rate Control With a Lasso Estimator. Submitted.

Tautenhahn R, Bottcher C, Neumann S. Highly sensitive feature detection for high resolution LC/MS. BMC Bioinformatics 2008;9:504.

Tautenhahn R, Patti GJ, Rinehart D, Siuzdak G. XCMS Online: A web-based platform to process untargeted metabolomic data. Analytical Chemistry 2012;84:5035-5039.

Tenenhaus M. La régression PLS. Revue de Statistiques Appliquées 1999;47:5-40.

Thevenot EA, Roux A, Xu Y, Ezan E, Junot C. Analysis of the human adult urinary metabolome variations with age, body mass index and gender by implementing a comprehensive workflow for univariate and OPLS statistical analyses. Journal of Proteome Research 2015;14:3322-3335.

Via A, Blicher T, Bongcam-Rudloff E, Brazas MD, Brooksbank C, Budd A, De Las Rivas J, Dreyer J, Fernandes PL, van Gelder C, Jacob J, Jimenez RC, Loveland J, Moran F, Mulder N, Nyrnen T, Rother K, Schneider MV, Attwood TK. Best practices in bioinformatics training for life scientists. Briefings in Bioinformatics 2013;14:528-537.

Weber RJM, Winder CL, Larcombe LD, Dunn WB, Viant MR. Training needs in metabolomics. Metabolomics 2015;11:784-786.

Wehrens R, Weingart G, Mattivi F. metaMS: An open-source pipeline for GC-MS-based untargeted metabolomics. Journal of Chromatography B 2014;966:109-116.

Weiss N, Barbier Saint Hilaire P, Colsch B, Isnard F, Attala S, Schaefer A, Amador MdM, Rudler M, Lamari F, Sedel F, Thabut D, Junot C. Cerebrospinal fluid metabolomics highlights dysregulation of energy metabolism in overt hepatic encephalopathy. Journal of Hepatology 2016;65:1120-1130.

Wishart DS, Tzur D, Knox C, Eisner R, Guo AC, Young N, Cheng D, Jewell K, Arndt D, Sawhney S, Fung C, Nikolai L, Lewis M, Coutouly MA, Forsythe I, Tang P, Shrivastava S, Jeroncic K, Stothard P, Amegbey G, Block D, Hau DD, Wagner J, Miniaci J, Clements M, Gebremedhin M, Guo N, Zhang Y, Duggan GE, MacInnis GD, Weljie AM, Dowlatabadi R, Bamforth F, Clive D, Greiner R, Li L, Marrie T, Sykes BD, Vogel HJ, Querengesser L. HMDB: the Human Metabolome Database. Nucleic Acids Research 2007;35:D521-526.

Xia J, Psychogios N, Young N, Wishart DS. MetaboAnalyst: a web server for metabolomic data analysis and interpretation. Nucleic Acids Research 2009;37:W652-W660.

# Supplementary material

**Create, run, share, publish, and reference your LC-MS, FIA-MS, GC-MS, and NMR data analysis workflows with the Workflow4Metabolomics 3.0 Galaxy online infrastructure for metabolomics**

Yann Guitton, Marie Tremblay-Franco, Gildas Le Corguillé, Jean-François Martin, Mélanie Pétéra, Pierrick Roger-Mele, Alexis Delabrière, Sophie Goulitquer, Misharl Monsoor, Christophe Duperier, Cécile Canlet, Rémi Servien, Patrick Tardivel, Christophe Caron, Franck Giacomoni, and Etienne A. Thévenot

Contents

# 1. Workflow Management with W4M

We provide here a brief description of the features offered by the Workflow4Metabolomics 3.0 online infrastructure (e-infrastructure). Some of these features, such as the building and sharing of workflows and histories, are provided by the Galaxy environment, which is used internally by W4M (Goecks et al., 2010). Other features are specific to the e-infrastructure, such as the computational tools themselves, the referencing of histories with DOI, the user accounts, the tutorials, the help desk, and the practical training sessions.

## 1.1. User Accounts

Accounts can be requested on the W4M home page. Accounts are private (i.e., only accessible to the user via his/her credentials).

## 1.2. Data Format

### 1.2.1. Raw data (for preprocessing tools)

#### 1.2.1.1. MS

MS data should be in either the mzML, mzXML, mzData, or NetCDF open format (Smith et al., 2006). We recommend the use of centroid data, in order to reduce file size. Conversion of raw data from proprietary format to centroid open format should be performed by the user before upload into W4M, e.g., by using the open-source ProteoWizard software (Chambers et al., 2012).

Raw files should be organized into a single or multiple folder(s), corresponding to the class(es) to be considered during preprocessing. The class information is used in the **xcms.group** tool, through the *minfrac* argument, to discard features which are not detected in a sufficient number of samples in at least one class (Smith et al., 2006). Separate classes may correspond to distinct sample types (blank reagent, quality control pools), or experimental conditions (treatment; Smith et al., 2006).

Raw files (either in a single or multiple folders) must be zipped before upload (we recommend the use of the 7-Zip open source software; http://www.7-zip.org).

For LC-MS, it is also possible now to upload the files individually (i.e., without folders and zip file). Within the history, the files must then be grouped as a data collection (Afgan et al., 2016) for further parallel processing by the **xcms.xcmsSet** tool. The use of a data collection therefore speeds up this computer intensive step. After peak detection, the collection of xset.RData outputs, together with a *sampleMetadata* file indicating the classes, are merged with the **xcms.xcmsSet Merger** tool before the grouping step (**xcms.group**). More details about the use of data collection for LC-MS data preprocessing can be found:

1) in the following tutorial:
http://download.workflow4metabolomics.org/docs/170510_galaxy_xcms_dataset_collection.m4v
2) and on the 'W4M_sacurine-subset_parallel-preprocessing' public history:
https://galaxy.workflow4metabolomics.org/history/list_published

### 1.2.1.2. NMR

NMR preprocessing tools currently work with Bruker files. Each sample directory should be organized with acquisition run and process numbered "1" (Table 1 and Fig. 8; upper right). Sample directories should then be gathered in a single parent directory, which should in turn be zipped before upload into W4M.

### 1.2.2. Preprocessed data (for normalization, quality control, statistical analysis, and annotation tools)

Preprocessing of the sample raw files generates a unique sample by variable data matrix of peak intensities, in addition to metadata of samples (e.g., sample ID, factor of interest) and variables (e.g., m/z and retention time). Such data and metadata are handled in W4M in a unique format consisting of 3 separate tables: *dataMatrix*, *sampleMetadata*, and *variableMetadata*. These tables, in a tabulated format (e.g., *.tsv*), are generated by the preprocessing tools from W4M, but can also be created or modified with spreadsheet editors (such as the Excel or the OpenOffice Calc software). The "3 table" format is used in all tools following the preprocessing (i.e., all tools for normalization, quality control, statistical analysis, and annotation). Details about the formats of the 3 tables can be found in the online tutorials (*HowTo* section from the front page). Once uploaded into W4M with the **Upload File** tool, the formats of the 3 tables can be verified with the **Check Format** tool (*Data Handling* section; note that the 'search tools' feature on top of the 'Tools' panel is helpful for locating a specific tool).

## 1.3. Data Upload

The **Upload File** tool allows data upload from a local computer into the W4M infrastructure. For sizes up to 2 GB (e.g., NMR raw zip or preprocessed data and metadata tables), files can be selected with a simple drag and drop. For bigger data sets (e.g., LC-MS and GC-MS raw zip), an FTP client software is required (such as Cyberduck, https://cyberduck.io, or WinSCP, https://winscp.net) to directly connect to the *ftp://workflow4metabolomics.org* infrastructure (with your W4M credentials) and copy/paste the zip file (see the *Galaxy Initiation* tutorial in the *HowTo* section).

## 1.4. Building Workflows

The user-friendly Galaxy features within the W4M e-infrastructure allow to build and save complex workflows. Workflows can be built *de novo* by using the *canvas* (editor): tools can be chained, and parameter values can be selected. Alternatively, workflows can be extracted from an existing history with the *Extract workflow* option (e.g., when the tools have been sequentially tuned on specific data).

## 1.5. Running Workflows

W4M offers users a high-performance environment for computing (4,000 cores) and data storage. The infrastructure can be accessed via a simple web browser and does not consume local resources. Once jobs are launched, the computation will continue and the results will be saved, even if the local connection is switched off by the user.

## 1.6. Sharing Histories

Histories can be shared with a dedicated user by using his/her W4M email (e.g., with colleagues within a lab or a consortium, or with a member from the help desk). Alternatively, histories can be published online to give all users unrestricted access to the workflow and the associated data (e.g., for training or citation purposes, see below).

## 1.7. Referencing Histories

Histories (workflow and all associated input and output data and metadata) can now be published on W4M 3.0 with a reference ID and a permanent Digital Object Identifier (DOI; see the *Result* section). Reference histories can then be cited in publications (see for instance Rinaudo et al., 2016), thus giving reviewers and readers full access to the analysis (i.e., data, metadata, workflow, parameters, and results).

## 1.8. Tutorials

Tutorials for data preparation, upload, and analysis, in addition to history sharing and publishing, are available in the *HowTo* section from the front page (http://workflow4metabolomics.org).

## 1.9. Help desk

For any question regarding the features of the infrastructure, the help desk is available at support@workflow4metabolomics.org.

# 2. Case Studies workflows

For each reference workflow designed for the three case studies, the tools, parameter values, and critical points are described hereafter. Histories can be accessed via the corresponding DOI permanent link, and imported into the user account with *Import history* followed by *start using this history*. Workflows can then be extracted by selecting *Extract workflow* from the *History options* menu.

## 2.1 Table S1: *W4M00002_Sacurine-comprehensive* (DOI:10.15454/1.481114233733302E12).

Please note that the statistical steps are identical to the *W4M00001_Sacurine-statistics* workflow.

| Tool | Step | Parameters (non-default values) | Variables | Critical points |
|---|---|---|---|---|
| ***Upload File*** | Data upload (negative ionization mode; 2 batches; 24 blanks + 26 QCs + 184 samples = 234 files; mzML format; centroid; 17.6 GB) | | | File splitting between subfolders impacts ***xcms.group*** (*minfrac* parameter) |
| ***xcms.xcmsset*** | Peak detection within each file | centWave: ppm = 3, peakwidth = c(5,20), snthresh = 10, mzdiff = 0.01, prefilter = c(3,5000), noise = 1000  [Original matchedFilter parameters: step = 0.01, | | Check the Total Ion Chromatogram (TIC) and Base Peak Chromatogram (BPC) graphical outputs, displayed as pairwise comparisons between the classes; See also the '1.2.1.1 Data Format' for parallel processing of the files (as a 'data collection') |

| Tool | Step | Parameters | Variables | Critical points |
|---|---|---|---|---|
| | | fwhm = 4, snthresh = 3] | | |
| *xcms.group* | Peak grouping between samples | bw = 5, minfrac = 0.1, mzwid = 0.01 | | Check on the graphical output the relevance of the selected parameter values |

| Tool | Step | Parameters | Variables | Critical points |
|---|---|---|---|---|
| *xcms.retcor* | Alignment of retention times between samples | extra = 1, missing = 40 | | Check on the graphical output the number of peak groups used for alignment, and their even distribution along the time axis |
| *xcms.group* | Peak grouping between samples | bw = 5, minfrac = 0.1, mzwid = 0.01 | | |
| *xcms.fillPeaks* | Imputation of missing values | | 7,456 (100%) [4,667 (100%)] | |
| *CAMERA.annotate* | Annotating features which may belong to the same metabolite (isotopes, adducts, fragments) | num_digits = 4 polarity = negative | | Select the appropriate polarity |
| *Upload File* | Import of the *sampleMetadata* file which contains the *sampleType*, *injectionOrder*, and *batch* columns (to be used in the **Quality Metrics**, **Generic Filter**, and **Batch Correction** tools), osmolality values for the **Normalization** tool, and age, BMI, and gender values for the **Univariate**, **Multivariate**, and **Biosigner** statistical tools | | | |
| *Check Format* | Checking format of *dataMatrix*, *sampleMetadata*, and *variableMetadata* files | | | |

| Tool | Step | Parameters | Variables | Critical points |
|---|---|---|---|---|
| *Quality Metrics + Generic Filter* | Discarding variables with: 1.a) *blankMean_over_sampleMean* > 0.5 or 1.b) *pool_mean* < 0.1 Discarding *blank* samples | | 6,041 (81%) [3,732 (80%)] | |
| *Batch correction* | Correcting each variable for signal drift and batch effect | *all_loess_pool* | | Include enough injections of the pool sample (at least 5 per batch to use *loess* correction), in particular at the first and last positions; check on the graphic if a correction is necessary, and which kind of model (*linear* or *loess*) provides the best fit |
| *Quality Metrics + Generic Filter* | Discarding variables with pool_CV > 0.3 Discarding *pool* samples | | 3,120 (42%) [2,102 (45%)] | If dilutions of the pool sample are provided, the correlation with the dilution factor is computed and can be used as an additional quality metric |
| *Normalization* | Dividing all intensities of a sample by its osmolality value | | | |
| *Transformation* | | *log10* | | |
| *Univariate* | Hypothesis testing of non-zero Spearman correlation with age or BMI (respectively of difference of medians between genders with the | *fdr* = 0.05 | | |

| Tool | Step | Parameters | Variables | Critical points |
|---|---|---|---|---|
| | Wilcoxon test), including a correction for multiple testing | | | |

| Tool | Step | Parameters | Variables | Critical points |
|---|---|---|---|---|
| *Multivariate* | OPLS (respectively OPLS-DA) modeling of the age or BMI response (respectively of the gender response) | *Number of orthogonal components* = NA (i.e. automatically optimized by the algorithm) | | Check the diagnostics; in particular the pQ2 value obtained after random permutation of the labels should be <0.05 to avoid overfitting |
| *Biosigner* | Selection of significant features for building either a PLS-DA, Random Forest, or SVM model of the gender response | *Seed* = 123 | | Set a specific seed value (Advanced computational parameters) to reproduce an identical signature from the internal bootstrap procedure |
| *HMDB MS search* | Search the HMDB database by the ion mz values | *Molecular Weight Tolerance* +- = 0.001 *Molecular Species:* negative mode | | |
| *Kegg Compounds* | Search the KEGG database by the neutral masses | *Delta of mass* = 0.001 | | |

## 2.2 Table S2: *W4M00004_GCMS-Algae* (DOI:10.15454/1.4811272313071519E12).

| Tool | Step | Parameters (non-default values) | Variables | Critical points |
|---|---|---|---|---|
| ***Upload File*** | Data upload (12 .CDF files converted from the Agilent Chemstation .D format) | | | File splitting between subfolders impacts ***xcms.group*** (*minfrac* parameter) |
| ***metaMS.runGC*** | Preprocessing of GC-MS data | | | Too high *similarity_threshold* value might result in some compounds with low intensity being misaligned; check the quality of the pseudospectra with the *GCMS_EIC* graphical output |
| ***Golm Metabolome Database*** | Search the Golm Metabolome Database with spectra in the *.msp* format | | | |
| *For additional database search* | | | | |
| *Download metaMS result* | Download locally the *.msp* file for NIST search | | | |
| *NIST search on local computer* | Read the *.msp* file with your *mssearch* software (usually preinstalled on your GC-MS workstation; for additional information, see the tutorial in the HowTo section from the W4M portal) | | | |
| *Creation of a spectral database* | Open the *peakspectra.msp* file; keep Unknowns 2, 4 and 5 only, and set the compound names to Citric acid, Mannitol, and Ribitol, respectively; when saving, make sure to keep the *.msp* (or *.txt*) file extension | | | |
| *Data re-processing using the user spectral database* | | | | |
| ***Upload file*** | Upload the modified *.msp* spectra database file into Galaxy | | | |

| Tool | Step | Parameters | Variables | Critical points |
|---|---|---|---|---|
| *metaMS.runGC* | Preprocessing of GC-MS data | Choose the *Use personal database* option and select your modified *.msp* file | 52 pseudo-spectra | |
| *Normalization* | Dividing the intensities of each sample by the dry weight value | | | |
| *Multivariate* | PCA and PLS-DA | Scaling = pareto, transformation = Log10 | | Check the diagnostics; for PLS models, the pQ2 value obtained after random permutation of the labels should be <0.05 to avoid overfitting |

## 2.3 Table S3: *W4M00006_BPAMmusculus* ([DOI:10.15454/1.4821558812795176E12](DOI:10.15454/1.4821558812795176E12)).

| Tool | Step | Parameters (non-default values) | Variables | Critical points |
|---|---|---|---|---|
| ***Upload data*** | Data upload (24 samples; Bruker format; 7 MB) | | | File name ("1") of the acquisition run and of the process of interest impacts ***NMR Bucketing*** |
| ***NMR Bucketing*** | Spectra bucketing and integration | BucketWidth = 0.01<br>LeftBorder = 9.5<br>RightBorder = 0.8<br>ExclusionZones:<br>  Left = 5.1<br>  Right = 4.5 | 809 buckets | Exclusion zones depend on the biological matrix and on the solvents used to prepare the samples |
| ***NMR Normalization*** | Spectra normalization to total intensity | Normalization method = Total | | |
| ***Multivariate*** | PCA | Scaling = pareto | | |
| ***Multivariate*** | OPLS-DA modeling of the treatment response | *Number of orthogonal components* = NA (i.e. automatically optimized by the algorithm)<br>Scaling = pareto | | Check the diagnostics; in particular the pQ2 value obtained after random permutation of the labels should be <0.05 to avoid overfitting |
| ***Generic Filter*** | Discarding variables with Treatment_OPLSDA_VIP_pred < 0.8 | | 157 buckets | |
| ***Univariate*** | Hypothesis testing of difference of means between treatment doses with the Wilcoxon test and the False Discovery Rate correction for multiple testing | corrected *p*-value significance threshold = 0.05 | | |

| Tool | Step | Parameters | Variables | Critical points |
|---|---|---|---|---|
| **NMR Annotation** | Spectra annotation based on an in-house database (175 reference compounds) | ExclusionZones: Left = 5.1 Right = 4.5 shift = 0.01 | 39 identified metabolites | Input spectrum can be a pool of all biological samples; alternatively, one spectrum from each class can be annotated; Exclusion zones depend on the biological matrix and on the solvents used to prepare the samples Identified metabolites should be checked by an NMR analyst |

# 3. References

Afgan et al. The Galaxy platform for accessible, reproducible and collaborative biomedical analyses: 2016 update. Nucleic Acids Research 2016;44: W3-W10.

Chambers et al. A cross-platform toolkit for mass spectrometry and proteomics. Nature Biotechnology 2012;30:918-920.

Goecks et al. Galaxy: a comprehensive approach for supporting accessible, reproducible, and transparent computational research in the life sciences. Genome Biology 2010;11:R86.

Rinaudo et al. *biosigner*: A new method for the discovery of significant molecular signatures from omics data. Frontiers in Molecular Biosciences 2016;3.

Smith et al. XCMS: Processing mass spectrometry data for metabolite profiling using nonlinear peak alignment, matching, and identification. Analytical Chemistry 2006;78:779-787.